\documentclass[usenatbib]{mn2e} 
\usepackage{longtable} 
\usepackage{graphicx}

\newcommand{\eg}{{\textrm e.g.\,}}

\def\halpha{\mbox{H$\alpha$}\,\,}

\def\egs{\,erg~s$^{-1}$}      

\def\av{\mbox{$A_{_V}$}}

\def\subsun{\mbox{$_{\normalsize\odot}$}}

\def\gsim{\;\rlap{\lower 2.5pt\hbox{$\sim$}}\raise 1.5pt\hbox{$>$}\;}
\def\lsim{\;\rlap{\lower 2.5pt\hbox{$\sim$}}\raise 1.5pt\hbox{$<$}\;}
\def\la{\mathrel{\hbox{\rlap{\hbox{\lower4pt\hbox{$\sim$}}}\hbox{$<$}}}}
\def\ga{\mathrel{\hbox{\rlap{\hbox{\lower4pt\hbox{$\sim$}}}\hbox{$>$}}}}

\def\arcmin{\hbox{$^\prime$}}
\def\arcsec{\hbox{$^{\prime\prime}$}}

\def\fdg{\hbox{$.\!\!^\circ$}}
\def\farcm{\hbox{$.\mkern-4mu^\prime$}}
\def\farcs{\hbox{$.\!\!^{\prime\prime}$}}

\def\ubvri{\hbox{$U\!BV\!RI$}\,}            

\def\ar{Astron. Reports}
\def\aj{AJ}
\def\actaa{Acta Astron.}
\def\apj{ApJ}
\def\apjs{ApJS}
\def\aa{A\&A}
\def\aas{A\&AS}
\def\basi{Bull. Astron. Soc. India}
\def\ibvs{IBVS}
\def\mnras{MNRAS}
\def\pasp{PASP}
\def\pasj{PASJ}
\def\sci{Science}

\title[Multi-wavelength study of a young open cluster NGC 7419]
      {Multi-wavelength study of a young open cluster NGC 7419}

\author[Himali et al.]
   {Himali Joshi$^1$, Brijesh Kumar$^{1,2}$, K. P. Singh$^3$, Ram Sagar$^1$,  
    Saurabh Sharma$^1$, \newauthor J. C. Pandey$^1$ \\ \\
    $^{1}$ Aryabhatta Research Institute of Observational Sciences, 
           Manora Peak, Nainital 263 129, India\\
    $^{2}$ Departamento de F\'isica, Universidad de 
           Concepci\'on, Casilla 160-C, Concepci\'on, Chile\\
    $^{3}$ Tata Institute of Fundamental Research, Mumbai 400 005, India\\
   }

\begin{document}

\date{\today}


\maketitle

\label{firstpage}

\begin{abstract}
Using new \ubvri\halpha CCD photometric observations and the archival infrared
and X-ray data, we have carried out a multi-wavelength study of 
a Perseus arm young galactic star cluster NGC 7419. An age of $22.5\pm3.0$ 
Myr and a distance of $3230^{+330}_{-430}$ pc are derived for the cluster. 
Our photometric data indicates a higher value of color excess ratio 
$E(U-B)/E(B-V)$ than the normal one. There is an evidence for mass segregation  
in this dynamically relaxed cluster and in the range $1.4-8.6M\subsun$, the mass function
slope is in agreement with the Salpeter value. Excess emissions in 
near-infrared and \halpha support the existence of a young 
($\le 2$ Myr) stellar population of Herbig Ae/Be stars ($\geq 3.0 M_\odot$) 
indicating a second episode of star formation in the cluster region. 
Using XMM-Newton observations, we found several X-ray sources in the 
cluster region but none of the Herbig Ae/Be stars is detected in X-rays. 
We compare the distribution 
of upper limits for Herbig Ae/Be stars with the X-ray distribution functions 
of the T-Tauri and the Herbig Ae/Be stars from previous studies, and found 
that the X-ray emission level of these Herbig Ae/Be stars is not more than
$L_X \sim 5.2\times 10^{30}$ \egs,  which is not significantly higher than 
for the T-Tauri stars. Therefore, X-ray emission from Herbig Ae/Be stars could be the 
result of either unresolved companion stars or a process similar to T-Tauri stars. 
We report an extended X-ray emission from the 
cluster region NGC 7419, with a total X-ray luminosity estimate of 
$\rm{\sim 1.8\times10^{31}~erg~s^{-1}~arcmin^{-2}}$. If the 
extended emission is due to unresolved emission from the point sources then 
we estimate $\sim$288 T-Tauri stars in the cluster region each having X-ray luminosity 
$\rm{\sim 1.0\times10^{30}~erg~s^{-1}}$.
Investigation of dust attenuation and ${}^{12}$CO emission map of a square 
degree region around the cluster indicates the presence of a foreground dust 
cloud which is most likely associated with the local arm star forming region 
(Sh2-154). This cloud harbors uniformly distributed 
pre-main-sequence stars ($0.1-2.0M\subsun$), with no obvious trend of their 
distribution with either $(H-K)$ excess or \av. This suggests that the star 
formation in this cloud depend mostly upon the primordial fragmentation.
\end{abstract}



\section{Introduction} \label{sec:int}

NGC 7419 ($\rm RA_{J2000}=22^h 54^m 20^s$, 
${\rm DEC_{J2000}}=+60\degr 48\arcmin 54\arcsec$; $l=109\fdg13$, $b=1\fdg12$), 
is a moderately populated young and heavily reddened galactic star cluster 
in Cepheus with a large number of Be stars.   
The cluster contains high mass ($\geq10M\subsun$); 
intermediate mass ($2-10M\subsun$)
and low-mass ($\leq2M\subsun$) stars. It is therefore an ideal 
laboratory for the study of initial stellar mass distribution as well as 
duration of star formation process in a molecular cloud. Presence of 
statistically significant number of Herbig Ae/Be stars in the cluster makes 
it very attractive for understanding the formation of these stars and origin 
of various atmospheric activities like \halpha emission and X-ray emission 
in them. However, to address these questions in detail, one would like 
to know accurate distance and age parameters of the cluster NGC 7419, which 
is lacking despite a number of photometric and spectroscopic studies. 
This is mainly because of the fact that the cluster is heavily reddened in 
comparison to the nearby clusters situated at the similar distances, and 
suffers from variable reddening.

In order to determine cluster reddening reliably, accurate $UBV$ broadband 
photometry of early type stars is essential. A comparison of the photometries 
available in the literature indicates that $UB$ data, many have systematic 
calibration error. For example, \citet{beauchamp94} have 
mentioned that their color may have an offset of $\sim0.2$ mag due to the 
calibration problems in $U$ band. Their photometric observations have been 
carried out in the poor seeing ($2\farcs5-4\farcs0$) conditions. This will 
affect cluster photometric data particularly in the crowded regions. 

The main goals of present study are to determine the distance, age and its 
spread, and mass function (MF) of the cluster as accurate as possible. 
This will help us to understand the star formation history of the cluster, 
and to investigate the X-ray emission properties of Herbig Ae/Be stars. 
Deep optical \ubvri observations ($V\sim22.0$ mag), narrow band \halpha
photometric observations along with the Two Micron All Sky Survey (2MASS), 
Infrared All Sky Survey (IRAS), Midcourse Space experiment (MSX) and 
XMM-Newton archival data are used to understand the X-ray emission 
properties of Herbig Ae/Be stars, and the global scenario of star 
formation in the cluster NGC 7419 and its surrounding region.
 
\citet{blanco55} has reported the distance of this cluster as $\sim$ 6 kpc 
based on the $RI$ photometric observations. A similar value for cluster 
distance has also been obtained by \citet{moffat73}. 
However, \citet{hulst54} has obtained a significantly 
smaller  distance of 3.3 kpc. Using CCD data, a distance of 2.0 kpc and 
2.3 kpc was estimated by \citet{bhatt93} and \citet{beauchamp94}, respectively.
The age estimated by \citet{bhatt93} is $\sim$ 40 Myr while \citet{beauchamp94}
have estimated a much younger age of $\sim$ 14 Myr. Recent CCD observations 
reported by \citet{subramaniam06} estimated its distance as 2.9 kpc and an 
age of 20--25 Myr.

The paper describes optical observations and the derivation of cluster 
parameters in \S\ref{sec:phoDat} and \S\ref{sec:proClu}. 
The near-infrared (NIR) data are dealt in \S\ref{sec:nirDat}, while
 distribution of young stellar objects (YSOs), MF and 
mass segregation are given in  \S\ref{sec:spaDis}, \S\ref{sec:masFun} 
and \S\ref{sec:masSeg}. Finally, the X-ray data and its analysis 
(for the first time) are 
described in \S\ref{sec:xraDat}, followed by the summary and 
conclusions in \S\ref{sec:sumCon}.


\section{Photometric Data} \label{sec:phoDat}

\subsection{Observations} \label{sec:obs}

\begin{table}
\caption{Journal of CCD observations of the cluster NGC 7419 and the 
  calibration region SA98 \citep{landolt92}.}             
\label{tab:optlog}      
\centering                          
\begin{tabular}{l c l }        
\hline\hline                 
           Date(UT)& Filter & Exposure Time (s)       \\          
                   &        & ($\times$no. of exposures)\\  
\hline                                                      
                   &  NGC 7419&                                                 \\
\hline 
         30 October 2005 & \halpha      & 300$\times$2, 120$\times$2, 1500$\times$3\\
                   & \halpha-cont & 300$\times$2, 120$\times$2,  900$\times$3\\
         07 November 2005 & $U$               & 300$\times$3, 360$\times$4\\
                   & $B$               & 180$\times$4, 240$\times$4\\
                   & $V$               & 180$\times$4, 180$\times$4\\
                   & $R$               & 20$\times$4,  120$\times$4\\
                   & $I$               & 20$\times$4,  60$\times$4\\
         08 November 2005 & $U$               & 1800$\times$1\\
                   & $B$               & 1200$\times$1\\
                   & $V$               & 900$\times$1\\
                   & $R$               & 300$\times$1\\
                   & $I$               & 500$\times$1\\
         25 October 2006 & $U$               & 300$\times$3\\
                   & $B$               & 300$\times$2\\
                   & $V$               & 100$\times$2\\ 
                   & $R$               & 20$\times$2\\
                   & $I$               & 20$\times$2\\
\hline                                                      
                   &  SA 98&                                                 \\
\hline 
         25 October 2006 & $U$               & 360$\times$11\\
                   & $B$               & 150$\times$10\\
                   & $V$               & 60$\times$10\\ 
                   & $R$               & 25$\times$10\\
                   & $I$               & 25$\times$10\\
       
\hline
   \end{tabular}
   \end{table}

The optical observations of NGC 7419 were carried out using 
a thinned back-illuminated CCD camera mounted at  
f/13 Cassegrain focus of the 104-cm Sampurnanand reflector telescope of 
Aryabhatta Research Institute of Observational Sciences, Nainital. 
A $24\mu$m square pixel of the $2048\times2048$ size CCD detector corresponds 
to 0\farcs38 and the entire chip covers a field of about 
$13\arcmin\times13\arcmin$ on the sky. In order to improve the signal 
to noise ratio, observations were taken in binned
mode of $2\times2$ pixel. The gain and readout noise of the CCD are 
10 electron per analog-to-digital unit and 5.3 electron, respectively.
The journal of optical observations is given in Table~\ref{tab:optlog}. 
Broad band Johnson $UBV$, Cousins $RI$ and, narrow band \halpha line 
($\lambda=656.5$ nm) and \halpha continuum ($\lambda=665$ nm) filters 
were used for observations. The narrow band filters had a full width at 
half maximum of 8 nm. Several bias and twilight flat field frames 
in all the filters were taken to clean the images.
Multiple long and short exposures were obtained for the cluster region. 
We observed 12 stars in \citet{landolt92} standards field (SA98)
covering a range in brightness ($11.93<V<15.90$) as well as in 
color ($0.157<(B-V)<2.192$) for calibrating the cluster observations. The 
airmass range is covered from 1.1 to 2.0 for the \citet{landolt92} standards, 
which was used for extinction determinations.

The photometric CCD data were reduced
using the IRAF\footnote{IRAF -- Image Reduction and Analysis Facility is 
distributed by the National Optical Astronomy 
Observatories, which are operated by the Association of Universities for 
Research in Astronomy, Inc., under cooperative agreement with the National 
Science Foundation (http://iraf.noao.edu).}
and ESO MIDAS\footnote{MIDAS -- Munich Image Data Analysis System 
is developed and maintained by ESO, the European Southern Observatory.} 
data reduction packages. Photometry of the bias-subtracted and flat-fielded CCD
frames was carried out using DAOPHOT-II software \citep{stetson87,stetson92}.
Magnitude of the stars obtained from different frames
were averaged separately for short and long exposures. When
brighter stars were saturated on deep exposure frames, their
magnitudes were taken only from unsaturated short exposure frames.
We used DAOGROW program for construction of an aperture growth curve 
required for determining the difference between aperture and profile 
fitting magnitudes.

\subsection{Calibrations}\label{sec:cal}

\begin{table}
\centering
\caption{The mean photometric errors ($\sigma$) in magnitude are given as a function of brightness range in the cluster region.}             
\label{tab:errors}      
\begin{tabular}{c c c c c c}        
\hline                 
Magnitude & $\sigma_U$ & $\sigma_B$ & $\sigma_V$ & $\sigma_R$ & $\sigma_I$ \\ 
range     &            &            &            &            &\\
\hline   
$< 14$          &   0.009    & 0.002     &  0.003    &  0.005    &  0.007\\
$14-15$           &   0.007    & 0.005     &  0.003    &  0.005    &  0.004\\
$15-16$           &   0.021    & 0.010     &  0.013    &  0.012    &  0.025\\
$16-17$           &   0.030    & 0.015     &  0.010    &  0.011    &  0.023\\
$17-18$           &   0.069    & 0.018     &  0.010    &  0.012    &  0.026\\
$18-19$           &   0.131    & 0.029     &  0.014    &  0.011    &  0.031\\
$19-20$           &            & 0.070     &  0.017    &  0.013    &  0.026\\
$20-21$           &            & 0.173     &  0.031    &  0.019    &  0.031\\
$21-22$           &            &           &  0.082    &  0.036    &  0.034\\
$22-23$           &            &           &  0.196    &  0.061    &  0.047\\
\hline
\end{tabular}
\end{table}

  \begin{figure}
  \centering
  \includegraphics[width=8cm]{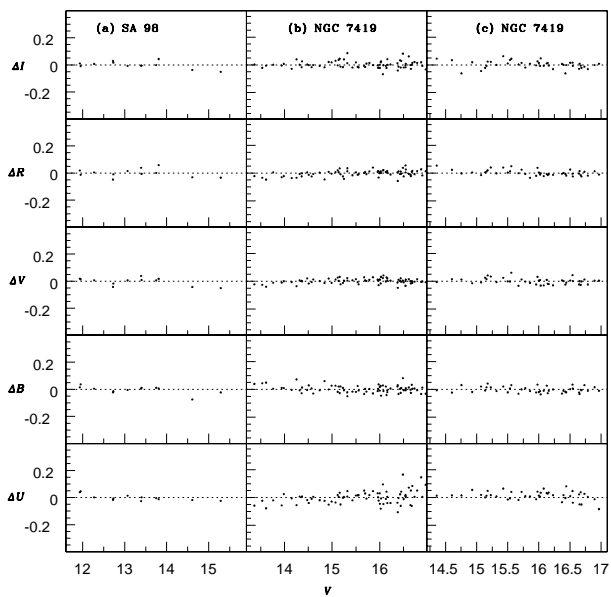}
  \caption{Difference in \ubvri magnitude measurements as a function of $V$ 
    magnitude are shown. The left panels (a) show the residuals of standard 
    magnitude from \citet{landolt92} and the present transformed magnitudes, 
    while the middle (b) and right (c) panels show 
    magnitude difference of secondary standards as generated on 25 October 
    2006 to that of on 07 November 2005 and 08 November 2005, respectively.}
  \label{fig:resd}
  \end{figure}

\begin{table*}
\tiny
\centering
\caption{$UBVRI$ and H$\alpha$ photometric data of the sample stars in 13$\arcmin$$\times$13$\arcmin$ region.
X and Y positions of stars in the CCD are converted into $\rm{RA_{J2000}}$ and 
$\rm{DEC_{J2000}}$ using the Guide Star Catalogue II (GSC 2.2, 2001). The complete table is
available in electronic form only.
}
\label{tab:datasample}
\begin{tabular} {c c c c c c c c c c c }
\hline
ID & $\rm{RA_{J2000}}$  & $\rm{DEC_{J2000}}$ & X & Y & $U$ &$B$&$V$&$R$&$I$&$\halpha$-index\\
   &(h m s) & (\degr~ \arcmin~ \arcsec) & (pixel) &(pixel) &(mag) &(mag)&(mag)&(mag)&(mag)&(mag)\\
\hline

        1   &   22 54 06.38 & +60 52 13.30&	  818.710 & 640.714 & 12.999  $\pm$  0.002  &12.481  $\pm$  0.002  &  11.707  $\pm$  0.009  & 11.321 $\pm$   0.012  & 11.003  $\pm$  0.002  &   0.032  $\pm$  0.008\\
        2   &   22 54 02.22 & +60 54 57.20&	 1032.410 & 691.088 & 13.518  $\pm$  0.002  &12.964  $\pm$  0.002  &  12.156  $\pm$  0.009  & 11.761 $\pm$   0.011  & 11.381  $\pm$  0.004  &   0.038  $\pm$  0.006\\
        3   &   22 53 30.17 & +60 53 38.60&	  914.148 & 994.003 & 13.964  $\pm$  0.003  &13.492  $\pm$  0.002  &  12.740  $\pm$  0.002  & 12.321 $\pm$   0.010  & 11.946  $\pm$  0.004  &   0.022  $\pm$  0.020\\

--&--&--&--&-- &--&-- &--& --&--&--\\ 
\hline
\end{tabular}
\end{table*}

The instrumental magnitudes were converted into the standard system 
using least square linear regression procedure outlined by \citet{stetson92}. 
The photometric calibration equations used are:\\
{\small \noindent
$u=U+(0.588\pm0.015)X+(-0.002\pm0.004)(U-B)+(6.933\pm0.006)$\\
$b=B+(0.347\pm0.013)X+(-0.035\pm0.005)(B-V)+(4.771\pm0.007)$\\
$v=V+(0.159\pm0.012)X+(-0.041\pm0.004)(V-I) +(4.320\pm0.007)$\\
$r=R+(0.139\pm0.009)X+(-0.056\pm0.007)(V-R)+(4.234\pm0.006)$\\
$i=I+(0.104\pm0.007)X+(-0.048\pm0.003)(V-I) +(4.767\pm0.005)$\\}

\noindent
where $U,B,V,R$ and $I$ are the standard magnitudes; $u,b,v,r$ and $i$
are the instrumental magnitudes obtained after time and aperture 
correction; and $X$ is the airmass. We have ignored the second-order 
color correction terms as they are generally small in comparison to 
the internal photometric errors as given by DAOPHOT. 
The photometric 
errors as a function of brightness range are given in Table~\ref{tab:errors}. 
It can be seen that the errors become large ($\geq0.1$mag) for stars fainter 
than $V\approx22$ mag, and hence the measurements beyond this magnitude are 
less reliable and are not considered in further analysis. At $V$ band, we 
could detect 1817 stars in $13\arcmin\times13\arcmin$ region and their photometric 
magnitudes are given in Table~\ref{tab:datasample} (available only in electronic form).
The difference
of calibrated magnitudes derived using above transformation to that of
the \citet{landolt92} magnitudes are plotted in Figure~\ref{fig:resd}.
We generated secondary standards in the cluster field using data 
of 25th October 2006 to standardise the data observed on 7th and 8th of 
November 2005. We, therefore also plot differences in magnitudes of the
secondary standards on these dates. No systematic effect has been seen in the 
residuals from night-to-night 
and its distribution is random in nature with a typical accuracy of $\sim0.03$ mag
in \ubvri band.

\begin{figure}
  \centering
  \includegraphics[width=8cm]{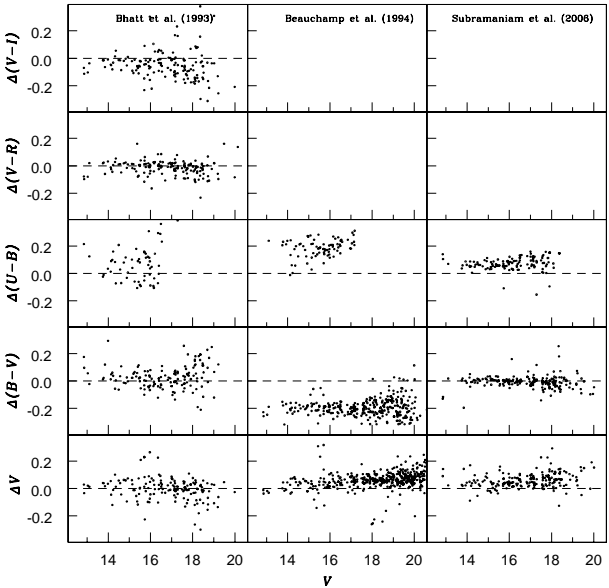}
  \caption{Comparison of present CCD photometry with that available 
    in the literature. The difference $\Delta$ (present-literature) 
    is in magnitude. The dashed line drawn in each panel represents 
    zero difference.}
  \label{fig:compare}
  \end{figure}

Figure~\ref{fig:compare} shows a comparison of the present CCD photometry  
with the previously reported CCD photometry by \citet{bhatt93},
\citet{beauchamp94} and \citet{subramaniam06}. The difference $\Delta$ 
(present-literature) is plotted as a function of $V$ magnitude and a 
detailed statistics are given in Table~\ref{tab:compare} (available only in electronic form).
In comparison to the photometric data of \citet{beauchamp94}, the $(B-V)$ 
colors obtained by us are bluer by $\sim0.25$ mag, $(U-B)$ colors are redder 
by $\sim0.25$ mag and the $V$ magnitude is fainter by $\sim0.05$ mag.
Our photometry is in agreement with \citet{subramaniam06} and 
\citet{bhatt93}. It is worth pointing out that our data is $\sim2.0$ mag 
deeper from \citet{subramaniam06} and \citet{bhatt93}, i.e., $V\sim$ 22.0 mag.

\begin{table*}
\centering                          
\caption{A comparison of present $UBVRI$ CCD photometry with that present in the
 literature -- viz \citet{bhatt93}, \citet{beauchamp94} and \citet{subramaniam06}.
The complete table is available in electronic form only.
}             
\label{tab:compare}      
\begin{tabular}{cccccc}        
\hline                 
  V range      &    $<\Delta V>$     &   $<\Delta (B-V)>$  & $<\Delta (U-B)>$ &$<\Delta (V-R)>$      &   $<\Delta (V-I)>$  \\
                      &    $Mean\pm \sigma$&   $Mean\pm \sigma$  & $Mean\pm \sigma$&     $Mean\pm \sigma$     &   $Mean\pm \sigma$  \\
\hline                 

\multicolumn{6}{l}{\citet{bhatt93}}\\
                    $< 14.0$      &            $0.025\pm 0.030$      (8) &             $0.060\pm 0.070$    (6)&            $0.030\pm 0.083$  (4)&   $-0.022\pm 0.065$  (7)& $-0.044\pm 0.034$  (6)  \\

--&--&--&--&--&--\\
\hline
   \end{tabular}
   \end{table*}

\subsection{Completeness of the data}\label{sec:comDat} 

 The completeness of the data
used in the present work was estimated using the ADDSTAR routine of the 
DAOPHOT II. In brief, the method consists of randomly adding artificial
stars (about 10--15\% of the originally detected stars) of known
magnitudes and positions into the original frames. The frames were re-reduced
using the same procedure used for the original frames. The ratio of the 
number of recovered stars to those added in each magnitude interval gives 
the completeness factor (CF) as a function of magnitude. The CF was obtained 
using the stars which were recovered in both $V$ and $I$ pass-bands. The 
detailed procedures have been outlined 
elsewhere \citep{sagar91,sagar98,pandey01,nilakshi02}. The CF as a function 
of $V$ magnitude is given in Table \ref{tab:cft}. 

\begin{table}
\caption{Completeness Factor of the photometric data as a function of 
  brightness and location.}             
\label{tab:cft}      
\centering                          
\begin{tabular}{c c c c}       
\hline                 
Magnitude range     &   \multicolumn{2}{c} { NGC 7419} & Field region \\
    ($V$ mag)      &   $r \le 1\arcmin$  & $1\arcmin \le r \le 3.5\arcmin$ & 
$3\farcm5 \le r \le 5\arcmin$ \\
\hline
13-14  &  1.00  &   1.00 &   1.00\\
14-15  &  1.00  &   1.00 &   1.00\\
15-16  &  1.00  &   1.00 &   1.00\\
16-17  &  1.00  &   0.98 &   0.99\\
17-18  &  0.89  &   0.97 &   0.98\\
18-19  &  0.57  &   0.93 &   0.96\\
19-20  &  0.59  &   0.89 &   0.97\\
20-21  &  0.60  &   0.86 &   0.93\\
21-22  &  0.15  &   0.38 &   0.47\\
\hline
\end{tabular}
\end{table}

\subsection{\halpha photometric data}\label{sec:halDat} 

\begin{table*}
\centering
\begin{minipage}{177mm}
\caption{Photometric properties of 44 \halpha emission-line stars in 
 $13\arcmin\times13\arcmin$ region around NGC 7419
 are tabulated. Columns 1-9 are taken from the present work. Column (10)
 carry comments on the stars identified in the present and previous
 work -- 'Yes' stands for detection and 'No' for non-detection in the present
 work, 'BMD' for \citet{beauchamp94}, 'PK' for \citet{pigulski00}, 
 'Sub' for \citet{subramaniam06} along with their identification in braces.  
 Stars located outside the cluster region 
 (see \S\ref{sec:halDat}) are indicated with 'field'.}
\label{tab:emstars}      
\begin{tabular} {c c c c c c c c c l}
\hline
ID& $\rm{RA_{J2000}}$  &$\rm{DEC_{J2000}}$& X      &  Y      & $V$   &  $R$    &$\alpha$-index  &$R-I$ &Remark     \\
  & (h  m  s)         &   (\degr~ \arcmin~ \arcsec) & (pixel)   &  (pixel)  &  (mag) &   (mag)  & (mag)   & (mag)& \\ 
(1)&(2)&(3)&(4)&(5)&(6)&(7)&(8)&(9)&(10)\\ \hline
  12  &  22 54 14  &    60 48 39 & 541.2 &549.5 & 13.75   & 12.63 & $-0.36$ &  $+1.08$  &  Yes, BMD(389), PK, Sub(M) \\ 
  15  &  22 54 39  &    60 47 24 & 454.5 &310.5 & 13.93   & 12.96 & $-0.01$ &  $+0.97$  &   No, BMD(1129), PK         \\
  18  &  22 54 26  &    60 49 02 & 576.0 &442.9 & 14.12   & 13.15 & $-0.10$ &  $+0.96$  &   No, BMD(781), PK, Sub(C) \\
  19  &  22 54 20  &    60 49 52 & 639.5 &496.0 & 14.14   & 13.05 & $-0.40$ &  $+1.05$  &  Yes, BMD(620), PK, Sub(G) \\
  20  &  22 54 15  &    60 49 50 & 634.7 &545.2 & 14.14   & 13.11 & $-0.18$ &  $+1.03$  &   No, BMD(417), PK, Sub(J) \\
  21  &  22 54 23  &    60 50 04 & 656.1 &473.7 & 14.20   & 13.07 & $-0.38$ &  $+1.10$  &  Yes, BMD(702), PK, Sub(4) \\
  23  &  22 54 23  &    60 49 53 & 642.3 &475.6 & 14.23   & 13.09 & $-0.56$ &  $+1.07$  &  Yes, BMD(692), PK, Sub(3) \\
  27  &  22 54 20  &    60 48 36 & 539.0 &499.1 & 14.36   & 13.33 & $-0.21$ &  $+1.02$  &   No, BMD(585), PK, Sub(I) \\
  29  &  22 54 37  &    60 48 36 & 547.6 &335.7 & 14.39   & 13.23 & $-0.50$ &  $+1.10$  &  Yes, BMD(1076), PK, Sub(A) \\
  38  &  22 54 27  &    60 48 52 & 564.0 &428.1 & 14.88   & 13.76 & $-0.54$ &  $+1.04$  &  Yes, BMD(831), PK, Sub(B) \\
  39  &  22 54 20  &    60 48 54 & 563.3 &492.1 & 14.93   & 13.92 & $-0.05$ &  $+1.04$  &   No, BMD(621), PK, Sub(K) \\
  48  &  22 54 18  &    60 48 57 & 566.2 &517.2 & 15.14   & 14.08 & $-0.50$ &  $+1.01$  &  Yes, BMD(504), PK, Sub(L) \\
  57  &  22 54 26  &    60 49 05 & 581.2 &435.0 & 15.37   & 14.30 & $-0.24$ &  $+0.81$  &   No, BMD(815), PK, Sub(2) \\
  59  &  22 54 08  &    60 50 23 & 673.6 &621.9 & 15.44   & 14.18 & $-0.55$ &  $+1.20$  &  Yes, BMD(239), PK, Sub(5) \\
  64  &  22 54 16  &    60 47 49 & 476.0 &532.4 & 15.54   & 14.32 & $-0.57$ &  $+1.16$  &  Yes, BMD(427), PK, Sub(N) \\
  66  &  22 54 14  &    60 46 20 & 358.0 &543.2 & 15.56   & 14.51 & $-0.22$ &  $+1.13$  &   No, BMD(375), Sub(P) \\
  69  &  22 54 24  &    60 49 31 & 613.6 &463.4 & 15.70   & 14.69 & $-0.44$ &  $+0.97$  &  Yes, BMD(728), PK, Sub(D) \\
  71  &  22 54 20  &    60 48 51 & 559.3 &500.3 & 15.71   & 14.76 & $-0.08$ &  $+0.96$  &   No, BMD(582), PK, Sub(H) \\
  73  &  22 54 32  &    60 47 57 & 494.2 &374.4 & 15.72   & 14.65 & $+0.04$ &  $+1.15$  &   No, BMD(967), PK         \\
  74  &  22 54 10  &    60 49 39 & 617.3 &595.2 & 15.73   & 14.67 & $+0.06$ &  $+0.59$  &   No, BMD(290), PK         \\
  75  &  22 54 29  &    60 49 08 & 586.1 &408.3 & 15.78   & 14.68 & $-0.52$ &  $+1.06$  &  Yes, BMD(884), PK, Sub(1) \\
  81  &  22 54 17  &    60 48 23 & 520.3 &526.5 & 15.95   & 14.98 & $-0.02$ &  $+0.99$  &   No, BMD(451), PK         \\
  85  &  22 54 17  &    60 48 18 & 514.4 &524.8 & 15.97   & 14.93 & $+0.05$ &  $+1.06$  &   No, BMD(458), PK         \\
  89  &  22 54 15  &    60 51 22 & 755.0 &556.0 & 16.01   & 14.92 & $-0.55$ &  $+1.06$  &  Yes, Sub(Q) \\
  91  &  22 54 13  &    60 50 09 & 658.0 &572.1 & 16.06   & 15.00 & $-0.01$ &  $+1.13$  &   No, BMD(351), PK         \\
  95  &  22 54 17  &    60 51 37 & 776.5 &535.6 & 16.11   & 15.10 & $-0.47$ &  $+1.01$  &  Yes, Sub(R)\\
 110  &  22 54 26  &    60 47 57 & 491.7 &435.1 & 16.35   & 15.28 & $-0.36$ &  $+1.04$  &  Yes, BMD(795), PK, Sub(6), \\
 126  &  22 54 12  &    60 48 33 & 532.6 &569.7 & 16.50   & 15.43 & $-0.01$ &  $+1.13$  &   No, BMD(340), PK          \\
 127  &  22 54 19  &    60 48 14 & 509.7 &506.9 & 16.50   & 15.45 & $-0.08$ &  $+1.11$  &   No, BMD(551), PK          \\
 139  &  22 54 24  &    60 47 36 & 462.8 &450.1 & 16.65   & 15.43 & $-0.51$ &  $+1.15$  &  Yes, BMD(745), Sub(E)  \\
 145  &  22 54 07  &    60 48 17 & 508.8 &618.9 & 16.71   & 15.33 & $-0.29$ &  $+1.33$  &  Yes                       \\
 181  &  22 54 51  &    60 48 27 & 543.4 &199.9 & 17.07   & 15.95 & $-0.28$ &  $+1.21$  &  Yes, field\\
 218  &  22 54 24  &    60 47 01 & 416.6 &447.5 & 17.32   & 16.17 & $-0.40$ &  $+1.12$  &  Yes, BMD(741), PK, Sub(F)  \\ 
 321  &  22 54 13  &    60 50 27 & 681.9 &572.7 & 18.13   & 16.92 & $-0.32$ &  $+1.23$  &  Yes, BMD(352),             \\
 438  &  22 54 38  &    60 52 10 & 829.5 &331.6 & 18.81   & 17.55 & $-0.30$ &  $+1.36$  &  Yes, field\\
 ---$^\dagger$  &  22 53 33  &    60 45 51 & 300.2 &938.4 & 18.87   & 17.27 & $-0.28$ &  $+2.11$  &  Yes, field\\
 655  &  22 54 56  &    60 47 41 & 485.2 &148.1 & 19.53   & 18.14 & $-0.46$ &  $+1.58$  &  Yes, field\\
 703  &  22 53 37  &    60 53 22 & 895.1 &925.5 & 19.65   & 18.45 & $-0.26$ &  $+1.24$  &  Yes, field\\
 714  &  22 54 40  &    60 49 24 & 612.6 &306.4 & 19.68   & 18.40 & $-0.35$ &  $+1.44$  &  Yes, BMD(1157)              \\ 
 716  &  22 54 57  &    60 53 38 & 954.5 &157.8 & 19.68   & 18.37 & $-0.32$ &  $+1.42$  &  Yes, field\\
 791  &  22 55 05  &    60 53 02 & 911.0 & 82.3 & 19.90   & 18.52 & $-0.43$ &  $+1.51$  &  Yes, field\\
 801  &  22 54 34  &    60 51 19 & 760.3 &366.9 & 19.92   & 18.59 & $-0.32$ &  $+1.47$  &  Yes \\
 808  &  22 54 40  &    60 48 31 & 542.9 &303.1 & 19.94   & 18.70 & $-0.26$ &  $+1.26$  &  Yes, BMD(1153)              \\
 831  &  22 54 13  &    60 45 59 & 329.3 &553.2 & 20.01   & 18.45 & $-0.48$ &  $+1.78$  &  Yes \\
\hline
\end{tabular}
\\
Notes: $^\dagger$ -- Error in Vmagnitude is greater than 0.1 mag.
\end{minipage}
\end{table*}

  \begin{figure}
  \includegraphics[width=8cm]{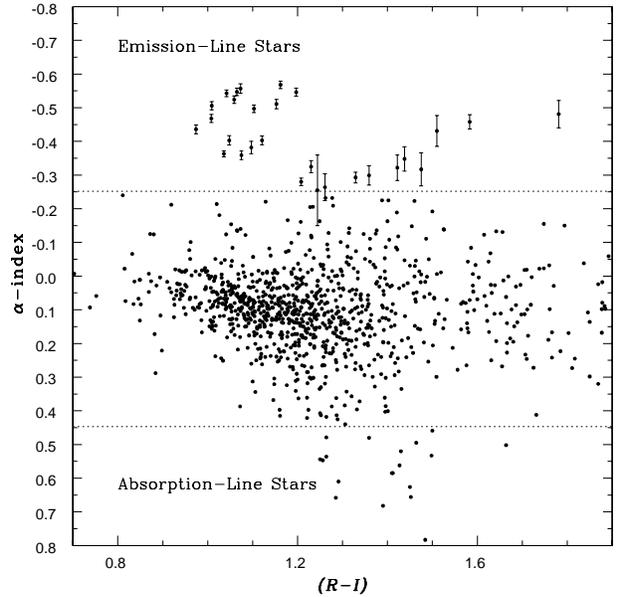}
  \caption{The $\alpha$-index as a function of color ($R-I$).
    The dashed line represents the $3\sigma$ limits from a mean $\alpha$-index 
    of $0.096\pm0.116$ mag. Emission-line stars having  $\alpha$-index  
    $\leq -0.252$ mag are shown with error bars. The absorption line stars 
    denote $\alpha$-index $\geq 0.444$ mag.}
  \label{fig:emstars}
  \end{figure}

To identify emission line stars, we use $\alpha$-index 
($m_{6565}-m_{6650}$) parameter, where $m_{6565}$ 
and $m_{6650}$ are the magnitude of a star in the \halpha and 
\halpha-continuum filter, respectively.  In an area 
of $13\arcmin\times13\arcmin$, we could measure $\alpha$-index for 1065 
stars and the same is plotted against $(R-I)$ colors in 
Figure~\ref{fig:emstars}. We estimated a mean $\alpha$-index iteratively for 
these stars by 3-sigma clipping. Seven such iterations resulted in a constant 
mean value of 0.097 mag and a RMS scatter ($\sigma$) of 0.116 mag. Stars 
having $\alpha$-index within $3\sigma$ are therefore considered as 
having normal strength at \halpha feature. A star is considered as 
\halpha emitter if $\alpha\leq-0.25$ mag and \halpha absorber if  
$\alpha\geq0.44$ mag. The observed characteristics of stars with 
\halpha emission are given in Table \ref{tab:emstars}. 
\halpha emission stars reported only by others in the literature are also 
listed in Table~\ref{tab:emstars}. This could be due to the variable nature 
of \halpha emission from the stars.

We detect 29 \halpha emission-line stars, and along with 15 more reported 
by \citet{pigulski00} and \citet{subramaniam06}, the total number of 
\halpha emission-line stars in the $13\arcmin\times13\arcmin$ region 
around the cluster have now increased to 44. Of these, thirteen are 
newly identified in the present work. Though some late-type dwarfs 
do show strong chromospheric \halpha in emission, but a few false 
detection may not be ruled out. As a non-emission late-type star 
generally has a series of strong metal oxide absorption lines, such as 
TiO in its  spectrum and this may lead to an under-estimation of their 
continuum, hence making the star an \halpha emitter \citep[e.g.][]{huang06}.

\section{Properties of the cluster}\label{sec:proClu}

\subsection {Radial density profile}\label{sec:radDen}

  \begin{figure}
  \includegraphics[width=8cm]{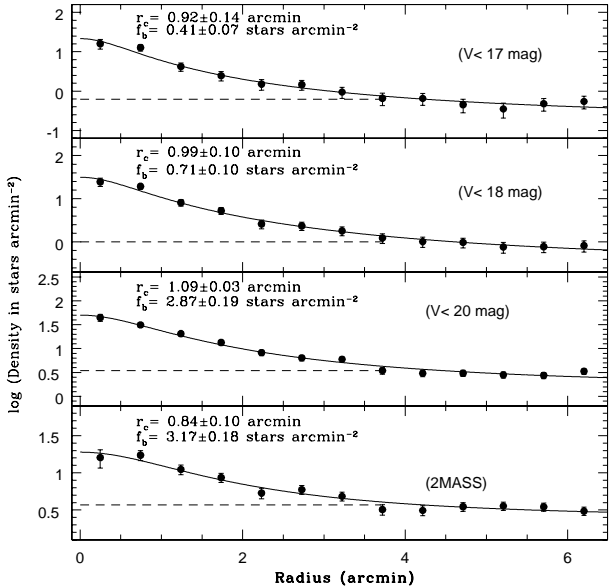}
  \caption{Projected radial stellar density profile of NGC 7419.
    Dashed lines represent 3$\sigma$ levels above the field star density 
    and solid curve shows best fit to the empirical model of 
    \citet{king62}.}
  \label{fig:rdp}
  \end{figure}
 
We estimate the cluster center iteratively by calculating average X and Y 
positions of stars with $V\leq18.0$ mag within 80 pixels from eye estimated 
center, until it converged to a constant value. The coordinates of the 
cluster center are found to be $\rm RA_{J2000} = 22^h 54^m 18^s$ 
and $\rm DEC_{J2000} =+60\degr 48\arcmin 58\arcsec$. Using this method, a 
typical error expected in locating the center is 5\arcsec.
To estimate extent of the cluster, we divide the cluster region into a number 
of concentric circles with an annulus of width $30\farcs5$. The projected 
radial stellar density profile is 
plotted in Figure~\ref{fig:rdp} for three magnitude limits -- 
\eg $V$=17, 18, 20 mag. The error bars denote Poisson statistics. 
The $\rho(r)$ is parameterized as \citep{king62,kaluzny92} 

  \begin{equation}
    \rho(r) \propto \frac { f_{\rm 0}} { 1+(r/r_{\rm c})^2}  
  \label{eqn:rdp}
  \end{equation} 

\noindent
where $r_{\rm c}$  and $f_{\rm 0}$ are the core radius of the cluster and
central star density, respectively. As the 
observed area ($13\arcmin\times13\arcmin$) is large compared to the
cluster diameter, we estimate $f_{\rm b}$ using outer region 
($r>5\arcmin$) of the cluster and to derive $f_{\rm 0}$ and $r_{\rm c}$, 
we performed a Levenberg-Marquardt non-linear fitting routine \citep{press92} 
to equation~\ref{eqn:rdp}. The fitted parameters are shown in 
Figure~\ref{fig:rdp}. We define extent of the cluster ($r_{\rm cl}$) as 
the radius at which $\rho(r)$ equals 3$\sigma$ level above the field star 
density. 
Using optical data, we have estimated the 
core radius of about 1\arcmin~and cluster radius of 4\farcm0$\pm$0\farcm5 
for the cluster. This is also supported by the radial density profile fits 
at 2MASS K-band data. The catalogue of open cluster by \citet{dias02} have 
been reported the radius of the cluster as 2\farcm5. 
To reduce the effect of field star contamination, we therefore, consider 
lower limit of radius as 3\farcm5 for further analysis.

\subsection{Color-magnitude diagrams}\label{sec:colMag}

  \begin{figure*}
  \includegraphics[width=15.0cm]{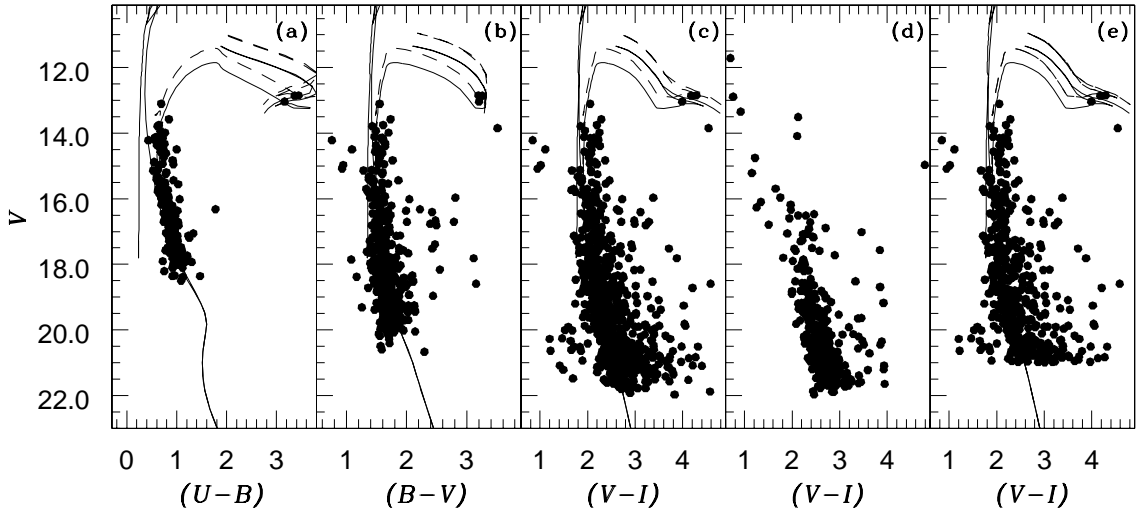}
  \caption{Color-magnitude diagrams for cluster regions are shown in 
    panels (a) $V,(U-B)$ (b) $V,(B-V)$ and (c) $V,(V-I) $. Panel (d) is for
    field region and the statistically cleaned $V,(V-I) $ CMD for cluster is
    shown in panel (e). Theoretical isochrones from GRD02 are 
    shown with continuous lines for log(age)= 6.6, 7.3 and 7.4 Myr in 
    the panels (a), (b), (c) and (e).}
  \label{fig:cmdAll}
  \end{figure*}

The $V,(U-B)$; $V,(B-V)$ and $V,(V-I) $ color-magnitude diagrams (CMDs)
of the cluster region along with $V,(V-I) $ CMD of the field 
region are shown in Figure~\ref{fig:cmdAll}. Morphology of the
CMDs is typical of a young-age open star cluster and the main sequence extends
down to $V\sim22.0$ mag in $V,(V-I)$ CMD. It is seen that the field star 
contamination becomes more evident for stars with $V>18$ mag, therefore,  
a statistical criteria was used to remove the field star contamination,
following procedure adopted by \citet{sandhu03}. For this purpose, we 
assume a uniform distribution of field stars in an area outside the 
cluster region ($r>3\farcm5$). The luminosity function 
(LF) derived from the $V,(V-I) $ CMD of the field region 
was subtracted from the LF of the cluster region. For a randomly selected 
star in the $V,(V-I) $ CMD of field region, the nearest star in the  
$V,(V-I) $ CMD of cluster region within $V~\pm0.25$ mag 
and $(V-I)~\pm0.125$ mag 
of the field star was removed. The necessary corrections for CF have also 
been considered in statistical subtraction. Statistical subtraction of data 
have been done till $V\approx21.0$ mag only as they have CF value $>$ 0.5.
The $V,(V-I) $ diagram of statistically cleaned cluster sample is shown in 
Figure~\ref{fig:cmdAll}. Further, to select main-sequence members more 
reliably, we define blue and red envelope in statistically cleaned CMD 
and the same is shown in the left panel of Figure~\ref{fig:cmdTau}.

\subsection{Interstellar reddening}\label{sec:intRed}

\subsubsection{Law of interstellar reddening}\label{sec:lawRed}

\begin{table*}
\caption{Observed color excess ratios in the direction of NGC 7419 are listed 
  along with the excess ratios valid for normal interstellr reddening law, 
  $R_V$=3.1 \citep{cardelli89}.}             
\label{tab:redlaw}      
\centering                          
\begin{tabular}{c c c c c c c c c c c}       
\hline                 
Object &$\frac{E(U-B)} {E(B-V)}$ &$\frac{ E(V-R)} {E(B-V)}$ &$\frac{ E(V-I)} {E(B-V)}$ &   $\frac{ E(U-B)} {E(V-J)}$ &$\frac{E(B-V)}{E(V-J)}$ &  $\frac{E(V-R)}{E(V-J)}$ &$\frac{E(V-I)}{E(V-J)}$ & $\frac{E(V-H)}{E(V-J)}$& $\frac{E(V-K)}{E(V-J)}$\\
\hline
NGC 7419 &0.90$\pm$0.09 & 0.62$\pm$0.02 &1.29$\pm$0.04 &0.37$\pm$0.04 & 0.41$\pm$0.01 & 0.25$\pm$0.01& 0.52$\pm$0.01& 1.14$\pm$0.01& 1.20$\pm$0.02\\
Normal value &0.72 & 0.62 & 1.25 & 0.32 & 0.43 & 0.27& 0.56& 1.13 & 1.21\\

\hline
\end{tabular}
\end{table*}

\begin{table*}
\caption{Giant stars with known spectral types in the 
         literature \citep{caron03,beauchamp94}. Color excess ratios 
         are derived using present data.}             
\label{tab:sptype}      
\centering                          
\begin{tabular}{c c c c c c c c c }       
\hline                 
ID  & BMD ID & Spectral Type& E(B-V) &  $\frac{E(U-B)}{E(B-V)}$ &  $ \frac{E(V-R)}{E(B-V)}$ &$ \frac{E(V-I)}{E(B-V)}$ &$ \frac{E(V-K)}{E(B-V)}$ & $R_V$=1.1$\frac{E(V-K)}{E(B-V)}$ \\ 
\hline
 8 &687  & B2.5 II  &  1.741   &   0.883  &   0.641  &   1.283 & 2.924 & 3.216\\
13 &473  & B2.5 III &  1.648   &   0.891  &   0.623  &   1.245 & 2.754 & 3.030\\
15 &1129 & B4.0 III &  1.669   &   0.752  &   0.639  &   1.266 & 2.833 & 3.116\\
16 &190  & B0.0 III &  1.999   &   0.924  &   0.642  &   1.267 & 2.837 & 3.120\\
20 &417  & B1/2 III &  1.845   &   0.912  &   0.638  &   1.263 & 2.903 & 3.193\\
12 &389  & B1/2 III &  1.789   &   0.896  &   0.670  &   1.313 & 3.014 & 3.315\\

\hline
\end{tabular}
\end{table*}

We investigated the nature of interstellar reddening towards the cluster 
direction using the color excess ratio method as described 
by \citet{johnson68}. We select stars with spectral types earlier than A0 by 
applying criterion of  
$V<17.7$ and $1.35<(B-V) <1.70$ on probable members of the cluster 
(\S\ref{sec:colMag}). From this list we exclude stars showing either \halpha emission 
(see\S\ref{sec:halDat}) or NIR excess (see\S\ref{sec:nirDat}) as their
reddening properties are likely to be different from the normal stars \citet{bkumar2004}. 
We estimate intrinsic colors using Q-method \citep[cf.][]{johnson53} and
iteratively estimate reddening free parameter 
Q [$=(U-B)-X(B-V)$, where $X=E(U-B)/E(B-V)$], 
till the color excess ratios become constant within the photometric errors. 
For the first iteration, we assume $X=0.72$. The color excesses are determined using 
intrinsic colors as derived from the MKK spectral type-luminosity class 
color relation given by \citet{caldwell93} for $(U-B)$, $(B-V)$, $(V-R)$ and $(V-I)$ 
and by \citet{koornneef83} for $(V-J)$, $(V-H)$ and $(V-K)$. Mean values of color 
excess ratios derived for the cluster are given in Table~\ref{tab:redlaw}
and except $X$, all other color excess ratios are in 
agreement with that expected for normal interstellar 
matter \citep{cardelli89}. The value of $X$ ($\sim0.90\pm0.08$) is 
significantly greater than the normal value of $\sim0.72$, which implies 
that the average dust grain sizes are smaller than average. 
The ratio of total-to-selective extinction is estimated using the 
relation ${R_V = 1.1E(V-K)/E(B-V)}$ \citep{whittet80} and it is found to 
be $3.2\pm0.1$.

We have also determined the value of color excess ratios using 
our $UBVRI$ photometry and spectral classification of six 
stars (see Table~\ref{tab:sptype}) available in the 
literature \citep{caron03,beauchamp94}. Intrinsic colors
\footnote{The intrinsic colors of Johnson $RI$  
in \citet{wegner94} are converted into Cousins $RI$ 
using \citet{bessell79} transformation relations.} 
are read from  \citet{wegner94} for $(V-R) $ and $(V-I) $, 
from \citet{fitzgerald70} for $(U-B)$. The mean value of  
$X$ ($=0.88\pm0.06$) using this method is in good agreement 
with the one derived from iterative Q-method. Further, the excess ratios 
at other wavebands, and mean value of $R_V$ ($=3.16\pm0.10$) are also close
to the one derived using Q-method (see Table~\ref{tab:sptype}). 

Higher value of color excess ratio $X$, as found above, is also supported 
by the $V,(U-B)$ CMD (Figure~\ref{fig:cmdAll}), in which the 
isochrones \citep[][hereafter GRD02]{girardi02} are found to fit well with 
$X=0.90$. Such anomalous values of $X$ are not uncommon for the interstellar
matter in the Milkyway and a higher value of $X$($>0.72$) 
has also been reported along lines of sight of other star forming clusters, 
for example, 
NGC 869 \citep[$X=0.90$;][]{pandey03};
Markarian 50 \citep[$X=0.81$;][]{baume04} and
NGC 7510 \citep[$X=0.78$;][]{barbon96}
However, they report normal value for other color 
excesses in these clusters.   
For further analysis, we have therefore used the normal value of 
color excess ratios except for the value $E(U-B)/E(B-V)$, which is 
adopted as 0.90.

  \begin{figure}
  \includegraphics[width=8cm]{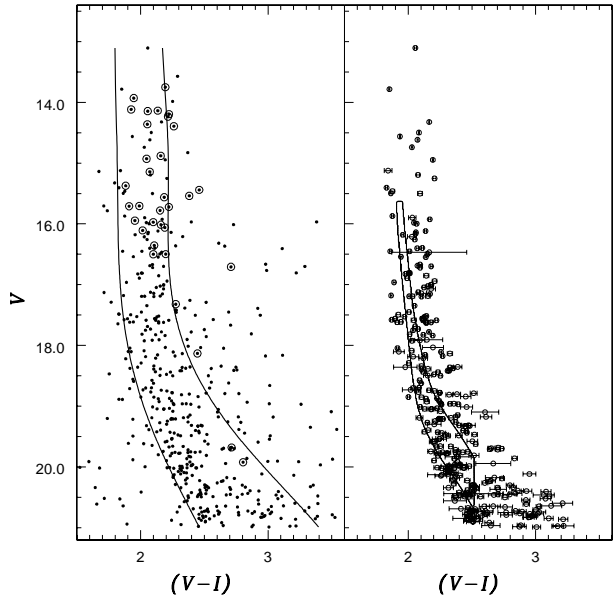}
  \caption{Statistically cleaned $V,(V-I) $ CMD for NGC 7419. 
    {\it Left panel:} Solid lines confine the blueward and redward boundary 
    of the main-sequence. Emission line stars are encircled. 
    {\it Right panel:} Solid curves show the best fit Padova isochrone 
    obtained using $\tau^{2}$-minimization method \citep{naylor06} for 
    fixed z=0.019 and age =25 Myr with distance as free parameter. 
    Horizontal bars denote errors in $(V-I) $. See \S\ref{sec:disClu} for 
    further details.}  
  \label{fig:cmdTau}
  \end{figure}

\subsubsection{Determination of reddening}\label{sec:detRed}

  \begin{figure}
  \includegraphics[width=8cm]{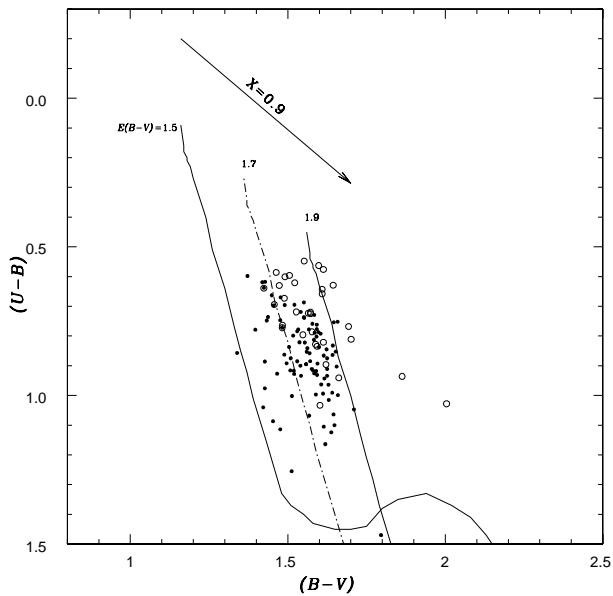}
  \caption{The $(B-V),(U-B)$ color-color diagram for stars in the cluster 
    region. Open circles represent stars with \halpha emission. Solid lines 
    are intrinsic main-sequence reddened along the reddening line with 
    $E(B-V)=1.5$ mag and 1.9 mag, while the dotted line is for 1.7 mag.}
  \label{fig:ccd}
  \end{figure}

Figure~\ref{fig:ccd} shows observed color-color, $(U-B),(B-V)$, diagram for the
cluster region. To determine reddening $E(B-V)$, the intrinsic colors for 
main-sequence stars with solar metallicity \citep{schmidtkaler82}  
are reddened along the reddening vector ($X=0.9$). This reddened main-sequence
suggests that $E(B-V)$ for the cluster region varies from 1.5 to 1.9 mag, with 
a visual mean value of about 1.7 mag. This reddening variation ($\sim0.4$ mag) 
for cluster members is well above the typical width of intrinsic main-sequence 
stars i.e. $\le 0.11$ mag, which is caused usually due to photometric 
accuracy and the presence of binary stars \citep{sagar87}. 
It therefore indicates the 
presence of variable reddening within the cluster region, which is also
reported by \citet{subramaniam06}. Few emission-line stars (open circles) 
are found to have $E(B-V)>1.9$ mag and these may either be embedded deep
in the parent cloud or be surrounded with circumstellar matter. Using 
iterative Q-method (\S\ref{sec:lawRed}), we derive mean value of  
${E(B-V)}$ as $1.7\pm0.2$ mag, which is same as estimated visually by
color-color diagram, and hence we adopt this value in our further analysis.

\subsection{Turn-off age of the cluster}\label{sec:turAge}

Turn-off age of the cluster is determined by comparing the theoretical 
stellar evolutionary models, the Padova isochrones GRD02 for 
solar metallicity Z=0.019 with observed CMDs (see Figure~\ref{fig:cmdAll}). 
Theoretical isochrones for ages 4, 20 and 25 Myr, corrected for the mean 
reddening $E(B-V)=1.7$ mag, are shown with solid lines and are visually 
fitted to the bluest envelope of CMDs consisting of probable members as 
selected in \S\ref{sec:colMag}. It is found that $X=0.9$ results in a better
isochrone fit in $V,(U-B)$ CMD than the normal value, $X=0.72$ 
(\S\ref{sec:lawRed}). Locations of the red supergiants \citep{caron03} 
BMD 921 (M2.5Iab), BMD 696 (M2.5Iab) and BMD 435 (M2.5Iab) are found to
be consistent with the isochrones, while BMD 139 (M3.5I) 
and BMD 950 (M7.5), both pulsating supergiants were prone to produce 
more scatter in distance and age estimate of NGC 7419. Thus, using
the visual-fit method, the present data finds the turn-off age of the 
cluster to lie between 19 Myr to 25 Myr and hence we assign an age 
of $22.5\pm2.5$ Myr.

We also estimate age using morphological age 
method \citep{phelps94,pandey97}, which employes color-index parameter (CIP),
defined as the difference in the color index between the blue turnoff 
point of the main-sequence and the color at the base of the red giant branch.
For NGC 7419, CIP is calculated as 1.654 mag and it represents the age between 
20 and 25 Myrs, which is in good agreement with our previous value. The 
present age estimate is similar to the one derived by \citet{subramaniam06} 
less than the 50 Myr given by \citet{bhatt93} and older than 
14 Myr derived by \citet{beauchamp94}.

\subsection{Distance to the cluster}\label{sec:disClu}

The zero-age main-sequence fitting procedure (Figure~\ref{fig:cmdAll}) 
is used to derive the distance of the cluster and the distance modulus 
is estimated as $12.45\pm0.20$ mag which corresponds to a distance 
of $3100\pm290$ pc. The distance determination is further performed using
a maximum likelihood $\tau^2$-minimization method which employes fitting 
two-dimensional distributions to stellar data in color magnitude 
space \citep{naylor06,mayne07,jeffries07}. This takes into account the 
effects of binary population as well as observed photometric errors. 
Assuming an age of $\sim22.5$ Myr and a binary fraction of 0.5, the best-fit 
Padova isochrone with solar metallicity yields a distance modulus 
of 12.55 mag (Figure~\ref{fig:cmdTau},\S\ref{sec:colMag}). Emission-line 
stars were excluded from this fit. Adding a systematic error of 0.28 mag due 
to differential reddening to each data point would result $P_r(\tau^2)=0.5$ 
with 68\% confidence ranges from 12.24 to 12.76 mag. We, therefore adopt
a distance of $3230^{+330}_{-430}$ pc for NGC 7419, which is similar to 
the estimate of \citet{subramaniam06} but higher from the value reported by
\citet{beauchamp94} and \citet{bhatt93}.

\section{Near-Infrared Data and Intermediate mass stars}
\label{sec:nirDat}

To understand the global scenario of star formation around NGC 7419, we 
use NIR $JHK_s$ data for point sources within a square degree 
region centered on $\rm RA_{J2000} = 22^h 52^m 28^s$ and 
$\rm DEC_{J2000} = +60\degr 53\arcmin 52\arcsec$, which 
is 14\farcm23 west from the cluster center. It contains the entire 
cluster and the surrounding star forming region. Data have been obtained 
from the 2MASS Point Source Catalog (PSC) \citep{cutri03}.
We have selected the sources based on the `read-flag' which gives the 
uncertainties in the magnitudes. We retain 2MASS sources with
`read flag' values of 1-2 for good quality data in our analysis.
The $JHK_s$ colors were transformed from 2MASS system to 
California Institute of Technology system using the relations given on 
the 2MASS web site
\footnote{http://www.astro.caltech.edu/$\sim$jmc/2mass/v3/transformations/} . 
  \begin{figure*}
  \includegraphics[width=8cm]{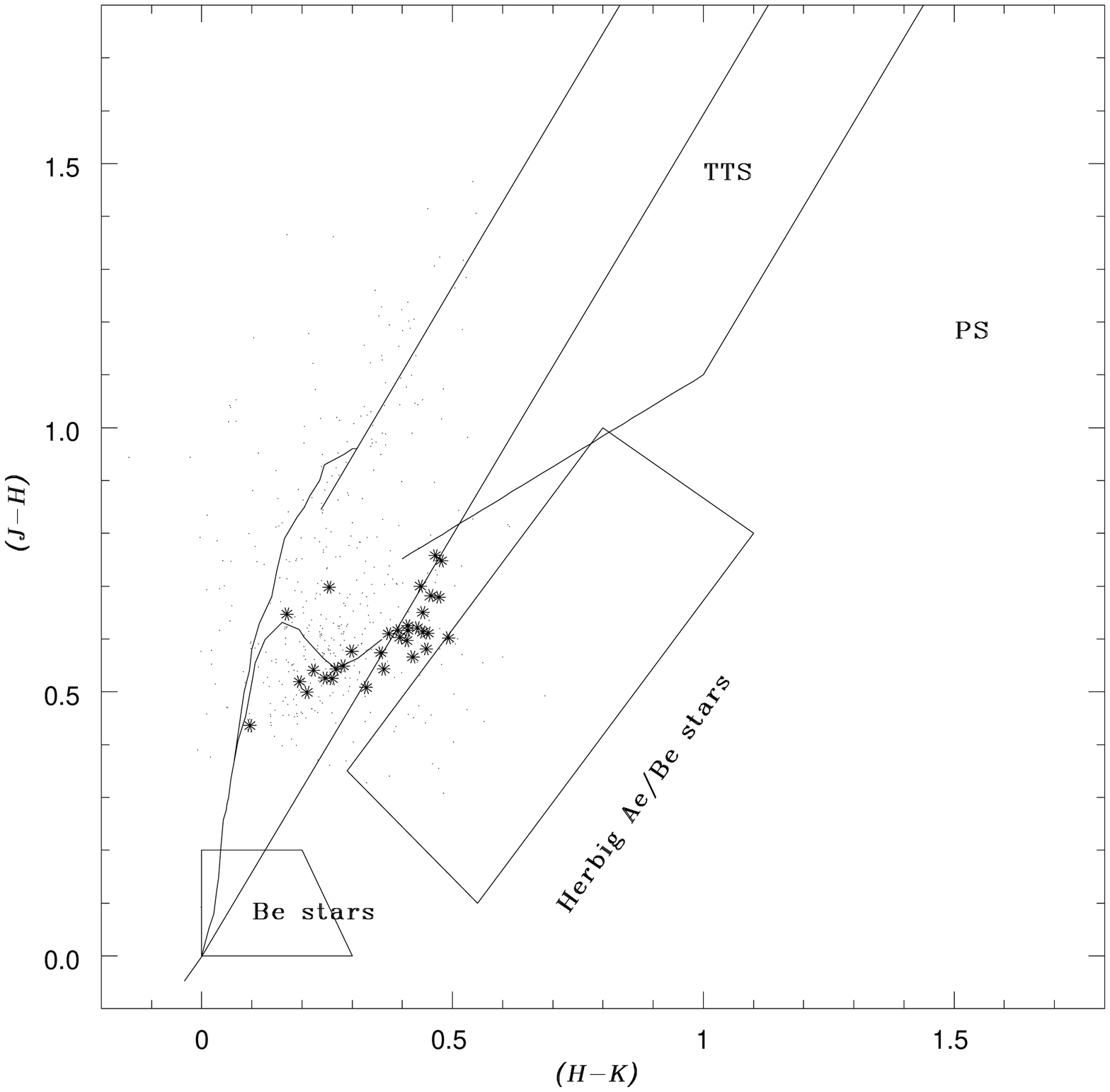} \hspace{4mm}
  \includegraphics[width=8cm]{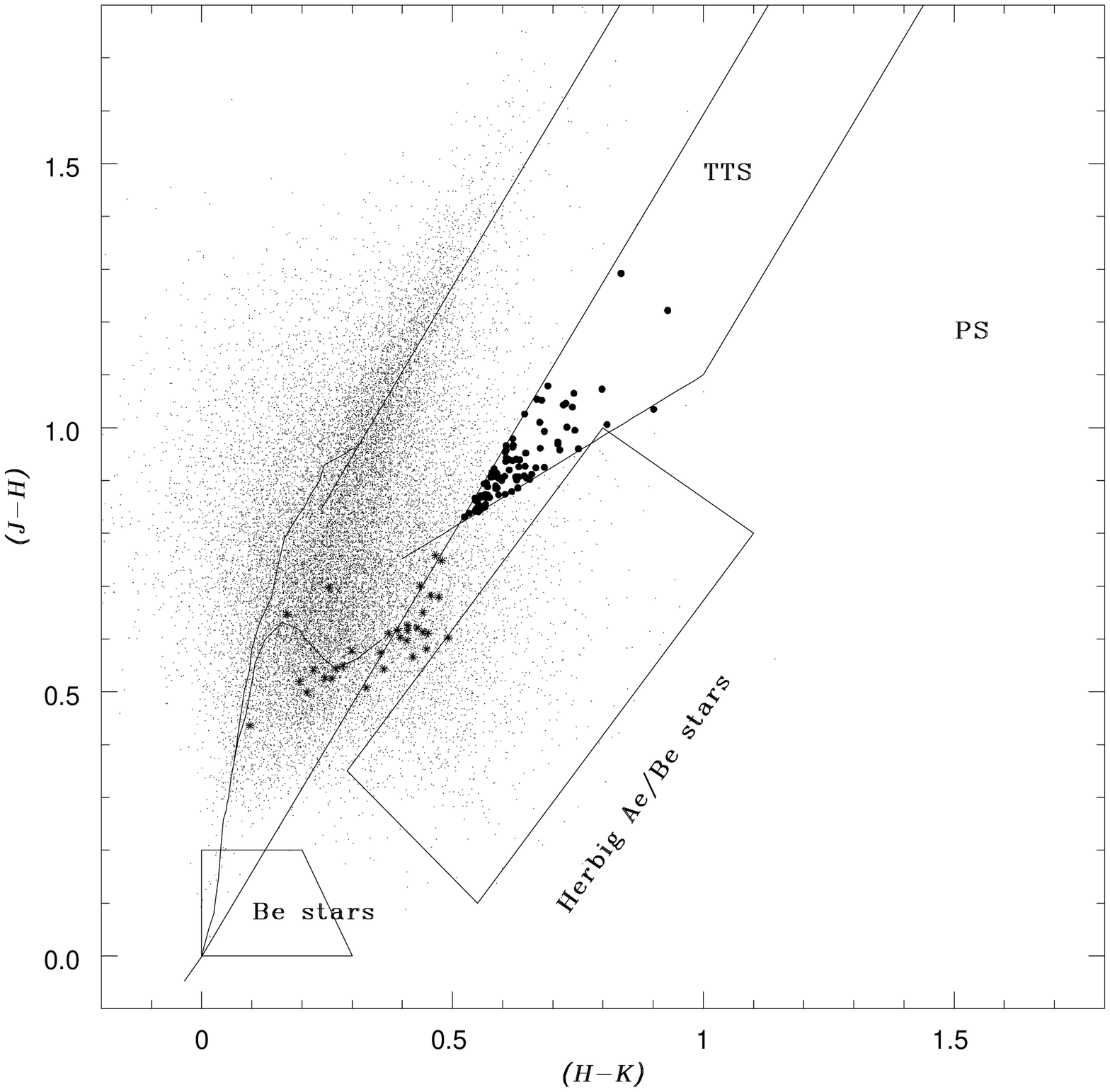}
  \caption{Color-color diagrams using the 2MASS JHK data. 
    {\it left panel:} The NIR color-color 
    diagram for the cluster region NGC 7419. Small dots represent the NIR 
    stars present in cluster region and the asterisks denote the \halpha
    emission-line stars from Table~\ref{tab:emstars}. 
    {\it right panel :} Same as the left panel, but for the square degree 
    surrounding region centered at $\rm RA_{J2000}= 22^h52^m28^s$ 
    and $\rm DEC_{J2000}= +60\degr 53\arcmin 52\arcsec$. 
    Bigger filled dots denote 90 T-Tauri stars in this region.}
  \label{fig:NIRccd}
  \end{figure*}

\subsection{Color-color diagrams}\label{sec:colCol}

The $(J-H),(H-K)$ color-color diagram is plotted for the cluster region 
(471 stars) and the surrounding square degree region (22100 stars), 
in Figures ~\ref{fig:NIRccd}. Solid lines represent unreddened 
main-sequence and giant branch \citep{bessell88}. The parallel 
solid lines are the reddening vectors for early and late-type stars 
(drawn from the base and the tip of two branches). Location of T-Tauri 
stars \citep[TTS;][]{meyer97}, proto-star (PS) like objects, 
classical Be stars and Herbig Ae/Be \citep{dougherty94,hernandez05} are 
also shown. The extinction ratio $A_J/A_V=0.282$, $A_H/A_V=0.180$
and $A_K/A_V=0.116$ have been taken from \citet{cardelli89}. 
Stars below the reddening vectors are considered to have NIR excess.
The asterisks represent all the stars with $\halpha$ 
emission (Table~\ref{tab:emstars}). Of 44 emission-line stars 
only 39 stars have NIR counterparts within $3\arcsec$. NIR excess is 
shown by 17 emission-line stars within the cluster region
but 4 more emission-line stars are situated  at the boundary of the 
reddening vector, therefore they have also considered as NIR excess stars 
within error bars in colors. From spectroscopic 
studies, \citet{subramaniam06} have identified them as Herbig Ae/Be stars. 
Therefore, NGC 7419 is a peculiar cluster having such a large  
number of Herbig Ae/Be stars. In surrounding region, 90 sources are 
found in the T-Tauri locus \citep{meyer97}. These sources are considered 
to be mostly classical T-Tauri stars (Class II objects) with large NIR 
excesses and belong to the YSO population. There may be an overlap in NIR 
colors of Herbig Ae/Be stars and T-Tauri stars in the TTS region 
\citep{hillenbrand92}. Such a large number of young stars represent 
the youth of this region. But all the T-Tauri stars with NIR excesses are 
situated outside the cluster region.

\subsection{Cluster Age from Herbig Ae/Be stars} \label{sec:cluAge}

\begin{figure}
  \includegraphics[width=8cm]{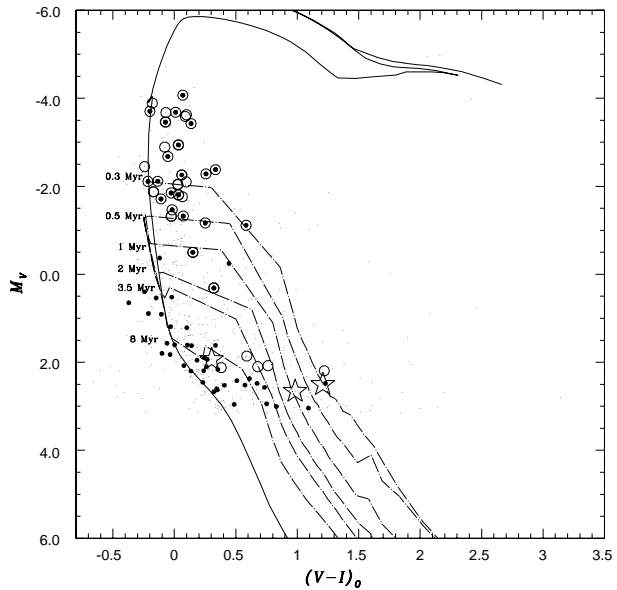}
  \caption{Color-magnitude diagram $M_V,(V-I) _0$ of the cluster region. 
    Open circles denote stars with $\halpha$ 
    emission (Table~\ref{tab:emstars}), asterisks denote stars with
    X-ray counterparts and having membership probability from X-ray 
    colors while filled dots represent stars with NIR excess. 
    Post-main-sequence isochrone for 25 Myr by GRD02 
    (continuous line) and pre-main-sequence isochrones for 0.3, 0.5, 
    1, 2, 4, 8 Myr by SES00 (dashed lines) are also shown. 
    Isochrones are corrected for a distance of 3.23 kpc.} 
  \label{fig:hebeage}
  \end{figure}

Figure~\ref{fig:hebeage} shows $M_V,(V-I) _0$ color magnitude diagram for 
\halpha emission stars (open circles; Table~\ref{tab:emstars}), NIR excess
stars (black dots), stars with X-ray 
counterpart (asterisks; \S\ref{sec:proMem}). Star BMD 950 is not shown as 
it shows a large error ($>0.1$ mag) in $I$ band. It is seen that most 
of Herbig Ae/Be stars are located to the right side of the main-sequence
and to estimate their age, we use pre-main-sequence isochrones 
from \citet{siess00}, hereafter SES00, for ages 0.3, 0.5, 1, 2, 4 and 8 Myr, shown with 
dashed lines in Figure~\ref{fig:hebeage}. For reference, the post-main-sequence 
isochrone of 25 Myr age GRD02 have 
been shown by continuous lines. Only one X-ray source and three emission-line 
star lie in the range of 3.5--8 Myr isochrones, which may arise due to
field star contamination of the Herbig Ae/Be stars loci in the NIR 
color-color diagram (Figure~\ref{fig:NIRccd}). Therefore, considering 
only the large number of Herbig Ae/Be stars, i.e. emission-line stars with 
NIR excess, it is very likely that the turn-on age of the cluster cannot 
be more than 2 Myr.

\subsection{Mass of Herbig Ae/Be stars using NIR color-magnitude diagram}
\label{sec:masHer}
  \begin{figure*}
  \includegraphics[width=8cm]{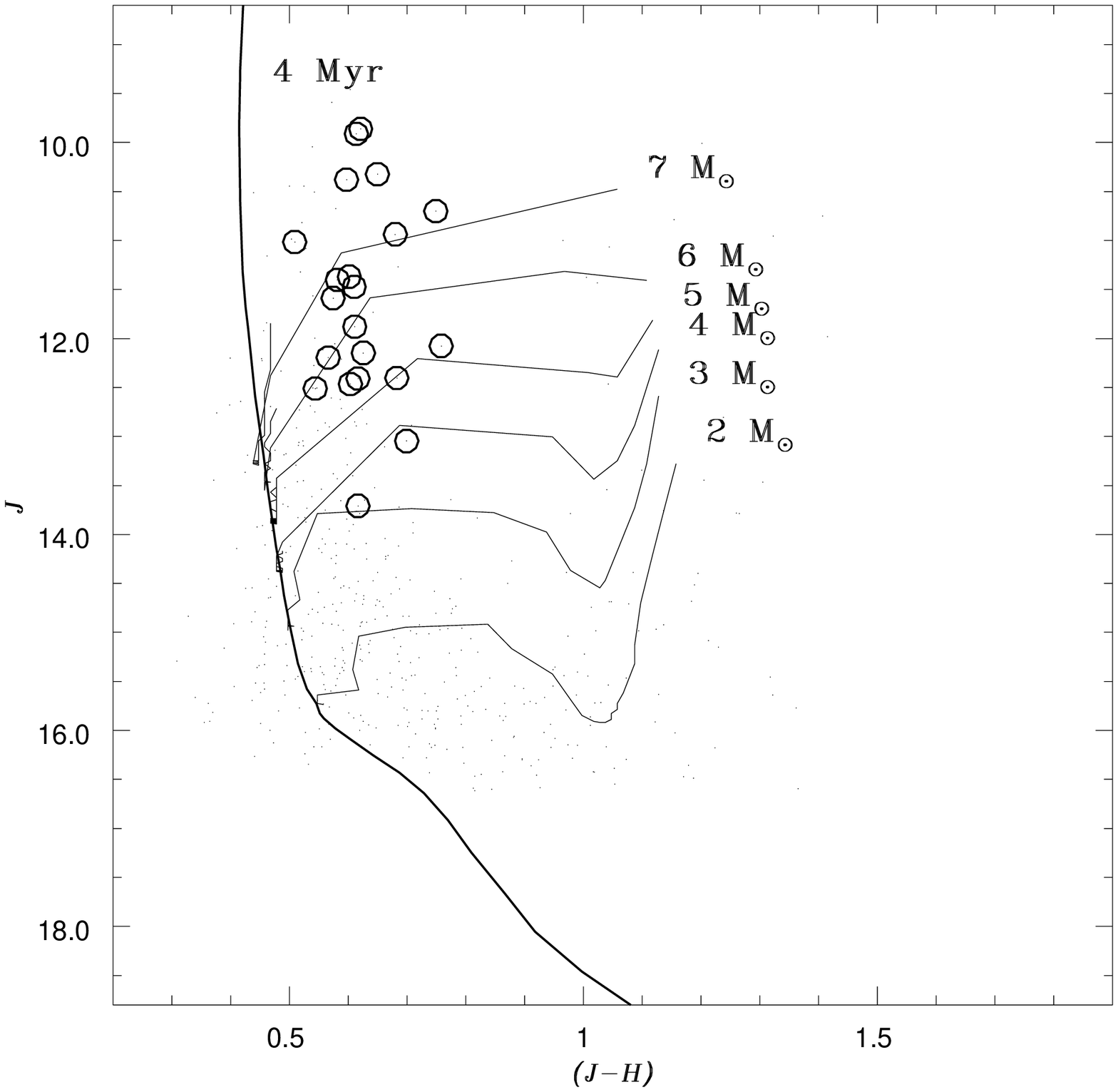} \hspace{4mm}
  \includegraphics[width=8cm]{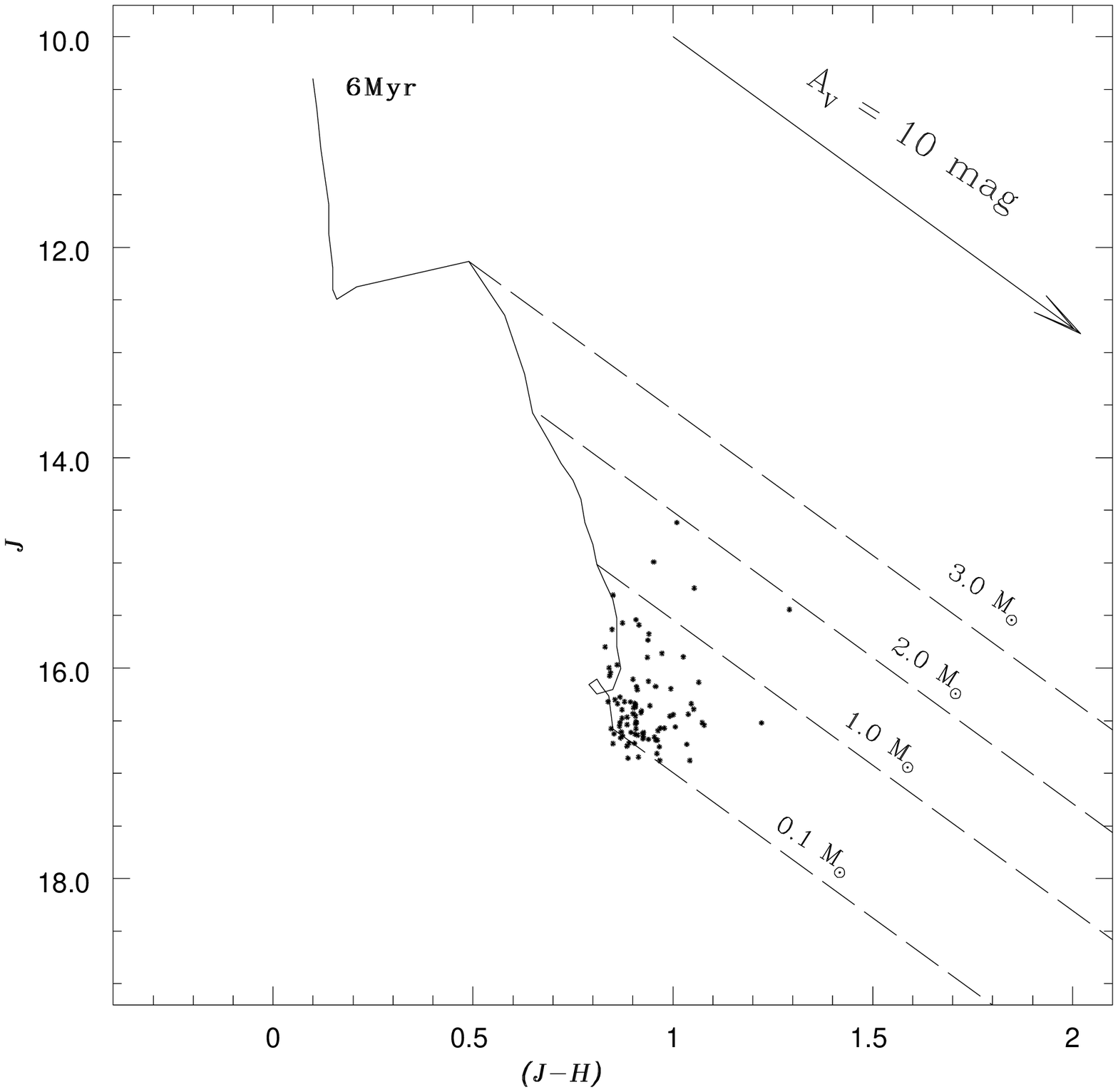}
  \caption{Stellar mass estimate using NIR CMD diagram. 
    {\it left panel:} $J, (J-H)$ CMD for Herbig Ae/Be stars in the cluster 
    region NGC 7419 are shown with open circles. The 4 Myr post-main-sequence 
    isochrone GRD02 and the pre-main-sequence evolutionary 
    tracks for masses $2-7 M\subsun$ are also shown and indicated.
    {\it right panel:} Same as left panel but for the YSO candidates 
    (asterisks) in the surrounding square degree region of NGC 7419. 
    Solid curve denotes pre-main-sequence isochrone of 6 Myr SES00,
    while the dashed oblique reddening lines denote the positions of 
    pre-main-sequence stars for masses 0.1, 1, 2 and 3$M\subsun$.}
  \label{fig:hebemass}
  \end{figure*}

Stellar masses are determined using NIR color-magnitude $J, (J-H)$ diagram. 
We prefer $J$ over $H$ or $K$, as the $J$ waveband is less affected by the 
emission from circumstellar material \citep{bertout88}. Left panel of 
Figure~\ref{fig:hebemass} shows NIR CMD for the cluster region NGC 7419. 
The 4 Myr post-main-sequence isochrone GRD02 and pre-main-sequence 
evolutionary tracks for mass range $2-7 M\subsun$ SES00 are plotted 
assuming a distance of 3.2 kpc and a mean reddening $E(B-V)$ of 1.70 mag 
(\S\ref{sec:intRed},\S\ref{sec:disClu}). Reddening and extinction corrections
to the isochrone and tracks are made using the relation  
$A_J/A_V=0.282$ and $A_{(J-H)}/A_V=0.102$ \citep{cardelli89}, where 
$A_V=3.1E(B-V)$. Location of 21 Herbig Ae/Be stars in NIR CMD is 
represented by open circles and it is seen that 66\% of these are located  
between the mass range $5-7M\subsun$. Only two stars lie in the mass 
range $3-4M\subsun$. 

Right panel of Figure~\ref{fig:hebemass} represents NIR CMD 
for 90 probable YSO candidates (shown with filled circles) identified
in a square degree region around the cluster (Figure~\ref{fig:NIRccd}). 
None of these YSOs are present in the cluster region and hence we argue  
that the population of YSOs is most likely associated with a local arm 
molecular cloud sh2-154 \citep{ungerechts00}, see \S\ref{sec:spaDis} for 
detailed discussion. We, therefore, correct the theoretical isochrones 
SES00, for extinction assuming the cloud distance 
of 1.4 kpc \citep{khalil04} and for reddening assuming a value of 0.4 mag/kpc
towards the direction of NGC 7419 \citep{joshi05}. The corrected
isochrones are shown with solid curves in the right panel 
of Figure~\ref{fig:hebemass} and they trace the locus of 6 Myr 
pre-main-sequence stars having masses in the range $0.1-3M\subsun$. 
It is observed that masses of all of the YSOs lie in the range  
$0.1-2.0M\subsun$ indicating them to be T-Tauri stars.
Therefore, almost all of the YSO candidates are low mass 
pre-main-sequence stars.

\section{Spatial distribution of low mass pre-main-sequence candidates}
\label{sec:spaDis}
\begin{table}
\scriptsize
\caption{IRAS point sources present in the square degree region centered 
  about $\rm RA_{J2000} = 22^h 53^m 26^s$ and
  $\rm DEC_{J2000} = +60\degr 53\arcmin 52\arcmin$. ID 44 is a carbon star.}     
\label{tab:iras}      
\centering                          
\begin{tabular}{c c c c c c c }       
\hline                 
ID & IRAS PSC  &  $F_{12}$ &  $F_{25}$ &  $F_{60}$ &  $F_{100}$ & Remark\\
   &         & (Jy) &   (Jy)&   (Jy)&   (Jy)&\\ 
\hline 
-- &22493+6018  &   7.04  &   2.39  &    3.12   &  48.91  &   VMW Cep\\ 
42 &22527+6030  &   2.53  &   0.77  &   17.05   &  87.48  &   VMZCep      \\
-- &22525+6033  & 112.10  &  92.30  &   17.05   &  50.10  &   M7.5Iab       \\
14 &22520+6031  &   3.10  &   1.06  &   17.05   &  50.87  &   M3.5 Iab     \\
-- &22512+6100  & 108.50  &  93.30  &   16.68   &  15.27  &   V386Cep            \\
-- &22538+6102  &   1.04  &   0.49  &    6.50   &  70.00  &   ---             \\
-- &22486+6028  &   0.85  &   0.37  &    4.95   &  53.07  &   ---            \\
-- &22490+6043  &   0.54  &   0.35  &    5.35   &  39.51  &   ---            \\
-- &22469+6053  &   0.46  &   0.25  &    5.55   &  71.76  &   ---            \\
-- &22480+6056  &   0.30  &   0.25  &    5.34   &  68.93  &   ---             \\
\hline
\end{tabular}
\end{table}

\begin{table}
\scriptsize
\caption{Same as for Table ~\ref{tab:iras} except for the MSX point sources.}             
\label{tab:msx}      
\centering                          
\begin{tabular}{c c c c c c c }       
\hline                 
ID&  Name     & $F_{A}$ &  $F_{B}$ &  $F_{C}$   & Remark\\
  &MSX PSC  &    (Jy)      &        (Jy)        &        (Jy)&   &\\ 
\hline                 
--  &G108.6966+01.0686  &          6.70    &        5.80    &        3.72    & VMWCep     \\ 
--  &G109.1599+01.1206  &         87.83    &      133.58    &       96.95    & M7.5 Iab     \\
42  &G109.1632+01.0645  &          3.31    &        2.14    &        1.48    & VMZ Cep\\
 5  &G109.1353+01.1118  &          3.40    &        2.00    &        1.68    & M2.5 Iab\\
 4  &G109.1302+01.1321  &          3.43    &        2.11    &        1.33    & M2.5 Iab\\
 7  &G109.1445+01.0947  &          1.71    &         0.71    &        0.74    & M2.5 Iab\\
14  &G109.0901+01.1184  &          3.38    &        2.41    &        1.58    & M3.5 Iab\\
--  &G109.2152+01.6007  &         63.48    &       83.70    &       66.06    & V386Cep\\
\hline
\end{tabular}
\end{table}

  \begin{figure}
  \vskip 6mm
  \includegraphics[width=8cm]{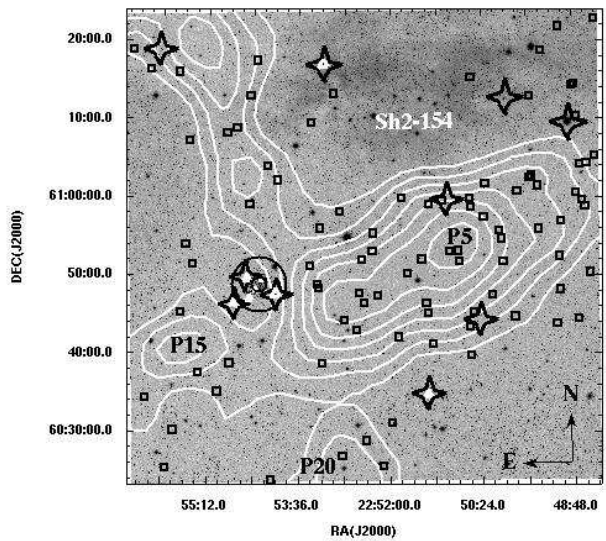} \hspace{4mm}
  \caption{A square degree $R$-band DSS image of the field surrounding 
    the cluster NGC 7419. North is up and east is to the left. 
    Open boxes show the candidate YSOs,
    white thick dots represent MSX sources and open star symbols denote 
    IRAS point sources. Inner ($r\le1\arcmin$) and outer boundaries 
    ($r\le3\farcm5$) of NGC 7419 are encircled. The location of Sharpless 
    nebula Sh2-154 is also marked.  
    The dust extinction maps produced using star counting 
    method are overlaid by white contours, and the associated clumps P5, P15 and P20 with cloud 699 
    are also marked \citep{dobashi05}.} 
  \label{fig:spaDis}
  \end{figure}

In Figure~\ref{fig:spaDis}, we show a square degree $R$-band image, reproduced 
from DSS\footnote{Digital sky survey -- http://stdatu.stsci.edu/dss/}, of 
the field containing YSO candidates and the cluster NGC 7419. Inner 
($r\le1\arcmin$) and outer boundaries ($r\le3\farcm5$) of the cluster
are encircled. Point sources from MSX (Midcourse Space Experiment) which
surveyed the Galactic plane in four mid-infrared bands - 
A(8.28 $\mu$m), C(12.13 $\mu$m), D(14.65 $\mu$m), and E(21.34 $\mu$m) 
at spatial resolution of $\sim18\arcsec$ \citep{price01}, are also shown 
in Figure~\ref{fig:spaDis}. Information about 10 IRAS (Infra-Red Astronomical
Satellite) point sources and 8 MSX point sources are given 
in Table~\ref{tab:iras} and Table~\ref{tab:msx}, respectively.
The MSX sources are denoted by white open circles while asterisks represent 
IRAS sources. YSO candidates are denoted by open square boxes. 
Of 8 MSX sources, six are identified with five supergiants and a carbon 
star and the other two are G108.6966+01.0686, a 
carbon star VMW Cep \citep{kukarkin68}, and G109.2152+01.6007, a semi regular 
pulsating star V386 Cep \citep{stephenson84}. The cluster supergiants also 
have IRAS counterpart, and their presence in NGC 7419 is also supported
from mid-infrared data. 

In the following, we describe the attenuation properties of a square degree region 
around the cluster (see \S\ref{sec:nirDat}). \citet{dobashi05} recently produced extinction maps of 
the entire region 
of the Galaxy in the galactic latitude range $b\leq 40\degr$ by applying 
traditional star count technique to the optical DSS sources. We retrieved 
the FITS images of the extinction map of the field containing NGC 7419 and 
their surrounding region from their on-line 
website\footnote{http://darkclouds.u-gakugei.ac.jp/astronomer/astronomer.html}. 
In Figure~{\ref{fig:spaDis}}, we overlay the contours of 
high resolution extinction map. The contours are plotted 
at $A_V=$ 2.5 to 4.5 with an interval of 0.3 mag. The extinction towards 
the location of the cluster shows relatively low ($A_V\le2.5$) values, 
in comparison to the outer region of the cluster especially, towards the 
west, the east and the south, where extinction increases up 
to $\sim4.5$, $\sim3.2$ and $\sim3.0$ mag, respectively. These 
enhanced attenuations are identified with the dense cloud clumps P5, P15 
and P20 of dark cloud 699 \citep{dobashi05}. Using star count method
it would be difficult to detect dust obscuration located at a cluster 
distance of $\sim$3.2 kpc, as the cloud would become inconspicuous due to 
large number of foreground stars \citep{dobashi05,medhi08}. We, therefore, 
argue that the dust obscuration would be due to a foreground cloud. 

\begin{figure*}
  \vskip 10pt 
  \includegraphics[width=7.8cm]{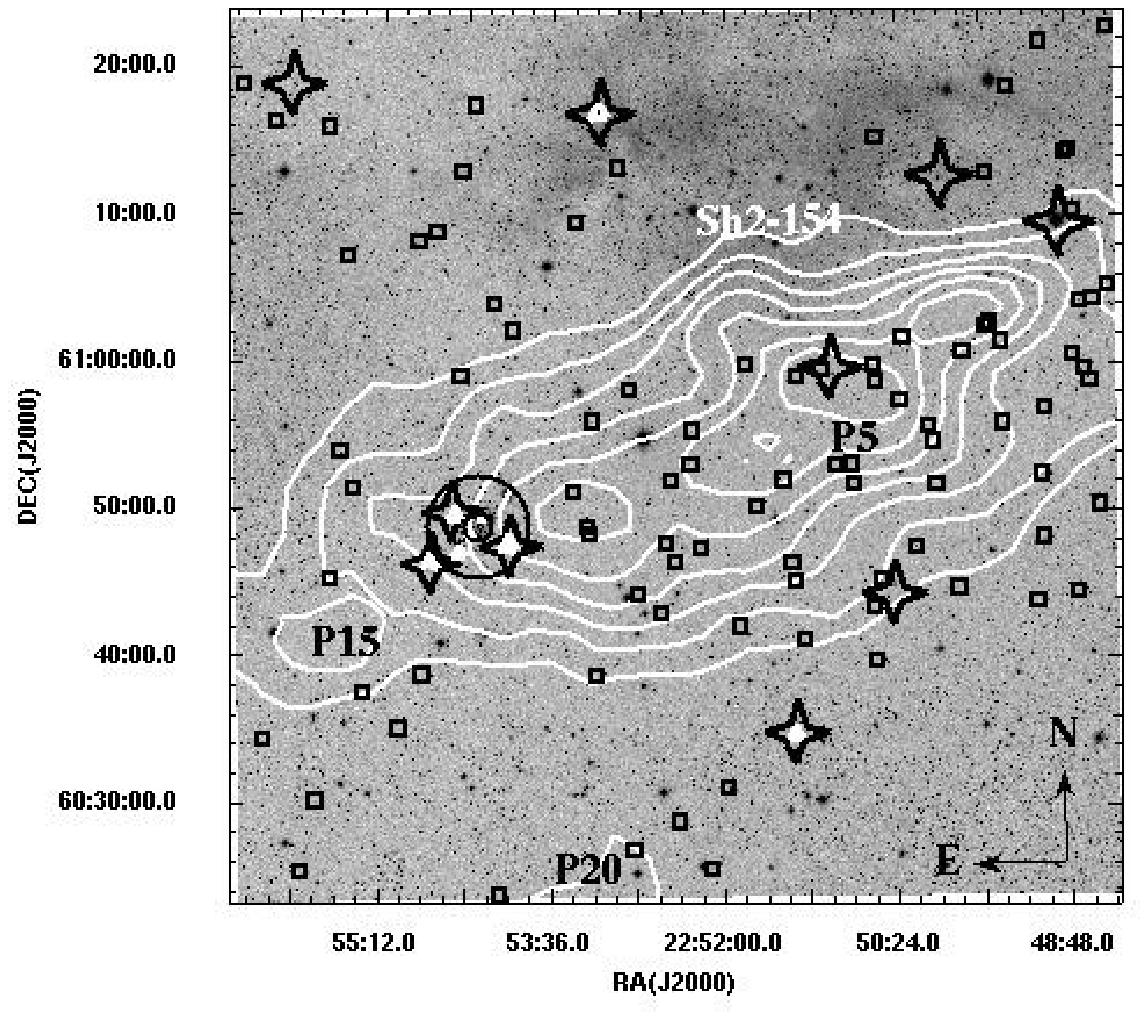} \hspace{47pt}
  \includegraphics[width=7.8cm]{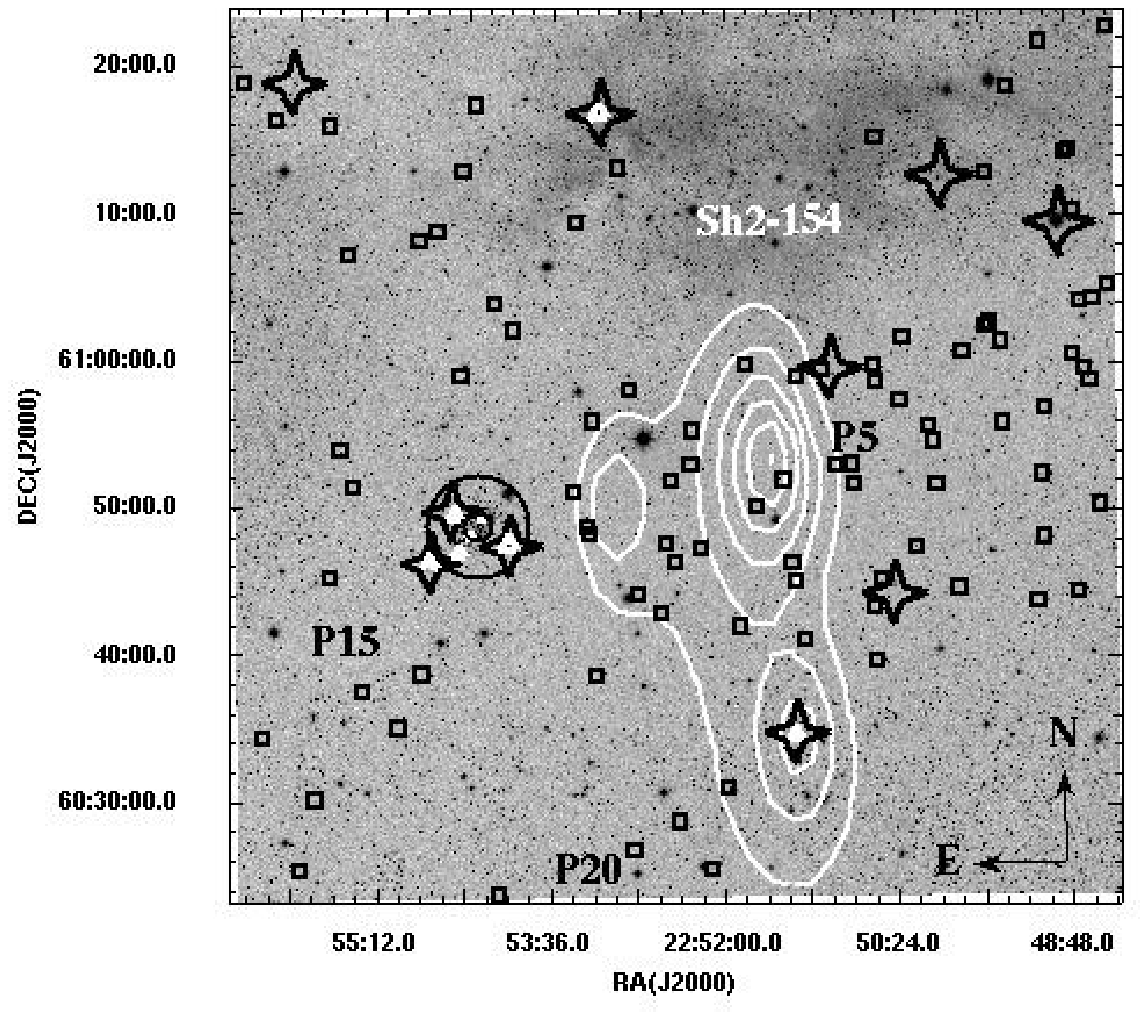}
  \caption{Same as Figure~\ref{fig:spaDis} but white counters represent in {\it left panel:} Contours 
    of $100~\mu$m IRAS dust emissions, in {\it right panel :}
    ${}^{12}$CO temperature map \citep{kerton03}.}
  \label{fig:spaDisr}
  \end{figure*}

In order to see the dust emission characteristics of this cloud, we 
superimpose the IRAS 100$~\mu$m contours obtained from the Infrared 
Processing and Analysis Center (IPAC) on $R$-band DSS image in 
Figure~\ref{fig:spaDisr}, left panel. The IRAS survey was done in four 
bands 12 $\mu$m, 25 $\mu$m, 60 $\mu$m and 100 $\mu$m from mid-infrared to far-infrared. 
The contours are plotted after smoothing the IRAS $100~\mu$m image. The contours are drawn 
at 134 (outermost) to 204 with the increment of 10 $\rm{ M~Jy~sr^{-1}}$.
The IRAS dust emissions support the presence of dust obscuration at the 
location of cloud clumps P5, P15 and P20. The star count extinction map is 
seen to be highly correlated with the 100$~\mu$m IRAS dust map and represent 
the same morphology of the cloud. We also show  the $^{12}$CO temperature map 
of the region, see right panel in Figure~\ref{fig:spaDisr} \citep{kerton03}.
The contours are drawn for temperature ranges $6.5-9.5$ K with the 
increment of 0.5 K. The peak of CO emission is most likely associated 
with clump P5 ($\sim5\farcm5$ westward). A slight mismatch might be the 
effect of low resolution of the extinction map \citep{dobashi05}. It is seen
that 60\% of total YSO population seems to be associated with the clumpy 
region P5. Therefore, the clump  P5 may provide a fertile environment for 
the formation of the low mass stars.

Interestingly, the spatial distribution of TT stars is highly correlated 
with the extinction and IRAS dust map (see Figure~\ref{fig:spaDis}), it
is therefore likely that the population of YSOs (\S\ref{sec:nirDat}) might be 
associated with the foreground dark cloud. \citet{avedisova02} identify
a local arm star forming region Sh2-154, situated in the north-east direction 
from the cluster. Optical diameter and distance of Sh2-154 is estimated 
respectively as $60\arcmin$ and 1.4 kpc \citep{blitz82}. Hence, the YSO 
population (Figure~\ref{fig:NIRccd}) could be part of Sh2-154 and associated 
with a dark cloud clump 699/P5 and 699/P15 as identified 
by \citet{dobashi05}.

\begin{figure*}
  \includegraphics[width=8.7cm]{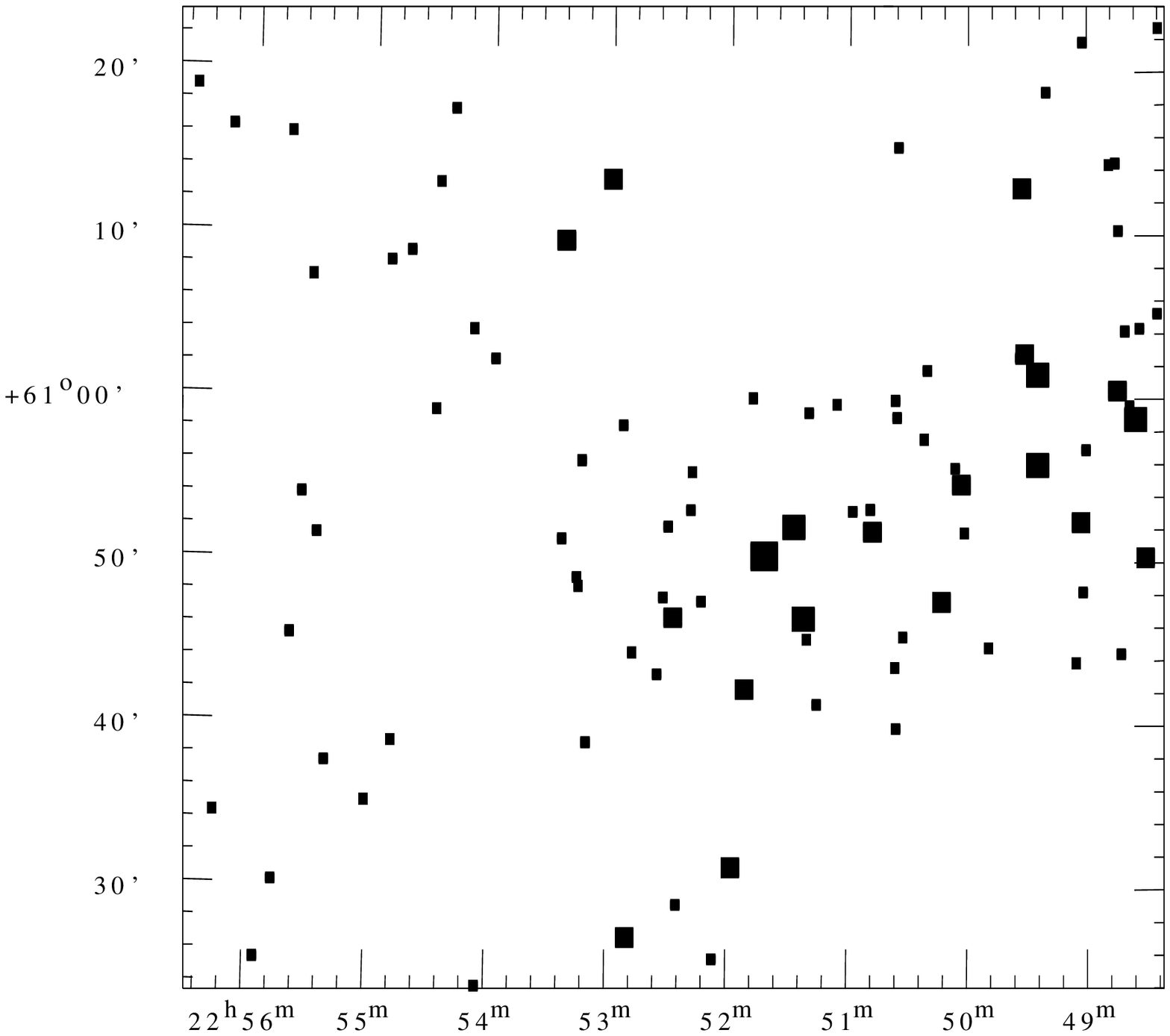} \hspace{4mm}
  \includegraphics[width=8.0cm]{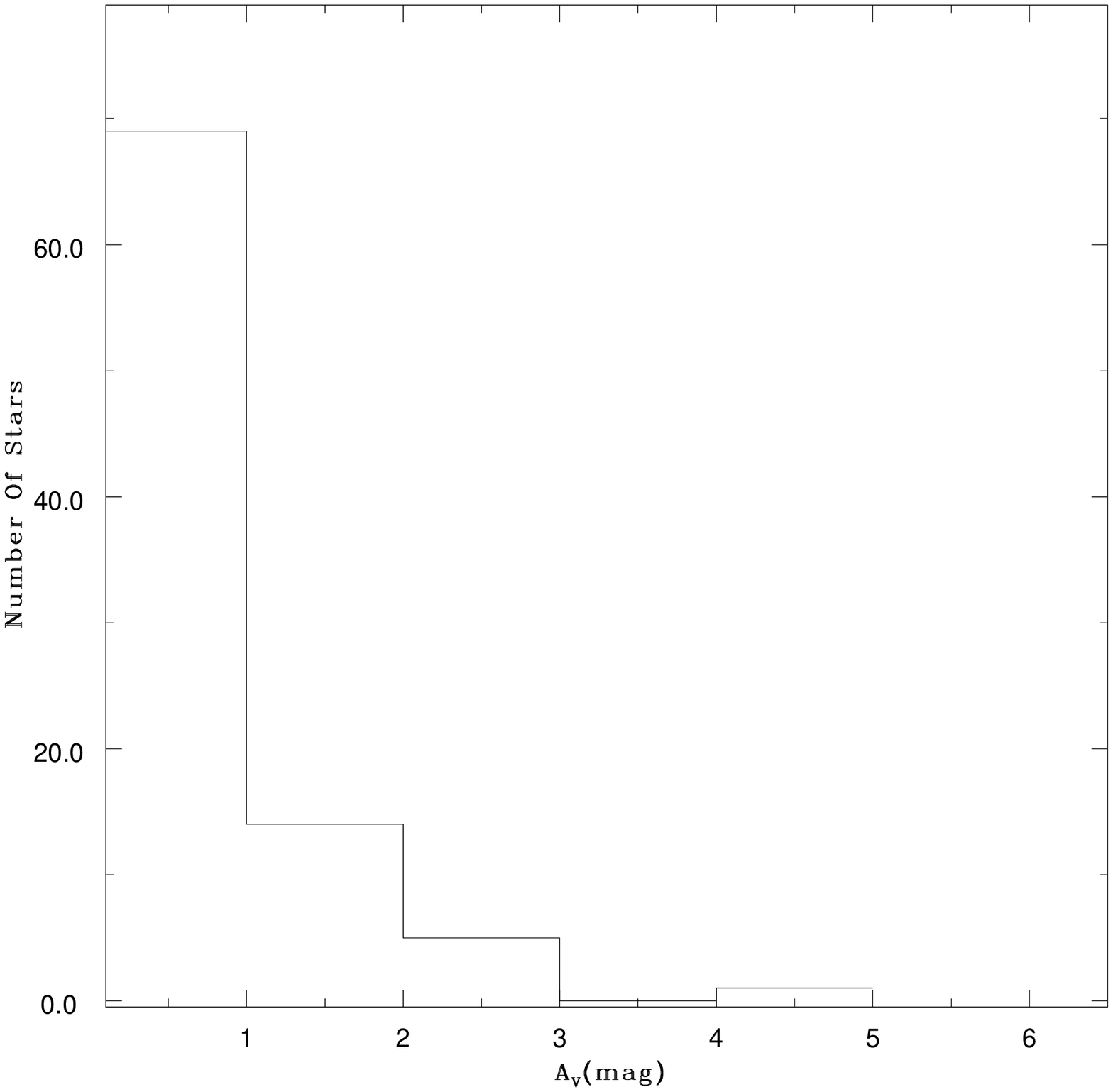}
  \caption{Distribution of the reddening. 
   {\it left panel :} Spatial distribution of the reddening of the YSO 
   candidates. Increasing sizes of filled boxes represent YSOs having $A_V$  
   with ranges $0-1$ mag, $1-2$ mag, $2-3$ mag and $4-5$ mag, respectively.
  X-axis and Y-axis denote $\rm{RA_{J2000}}$ and $\rm{DEC_{J2000}}$, respectively. 
   {\it right panel :} Frequency distribution of the reddening for YSO 
   candidates.}
  \label{fig:spaAv}
  \end{figure*}

The effect of environment on the production of the low mass stars
has been investigated with the help of the distribution of the reddening 
in the cloud by assuming that they are situated at the same distance.
The $A_V$ value for each star was measured by tracing back to the intrinsic 
lines along the reddening vector found in \citet{meyer97}. 
The value of $A_V$ is found to vary from 0 to 3 mag. Only one YSO 
has $A_V$ $\sim$5 mag. The spatial distribution of the $A_V$  
from 0 to 5 mag is shown in Figure~\ref{fig:spaAv}, left panel. 
No spatial gradient of $A_V$ is found with the spatial distribution of 
the YSOs. The frequency distribution of $A_V$ 
(Figure~\ref{fig:spaAv}, right panel) represents that
77\% YSOs are having $A_V$ less than 1.0 mag and 15\% of YSOs with
$A_V$ in between 1.0 to 2.0 mag. It shows that the environment for the 
YSOs is almost similar.

\begin{figure*}
  \includegraphics[width=8.7cm]{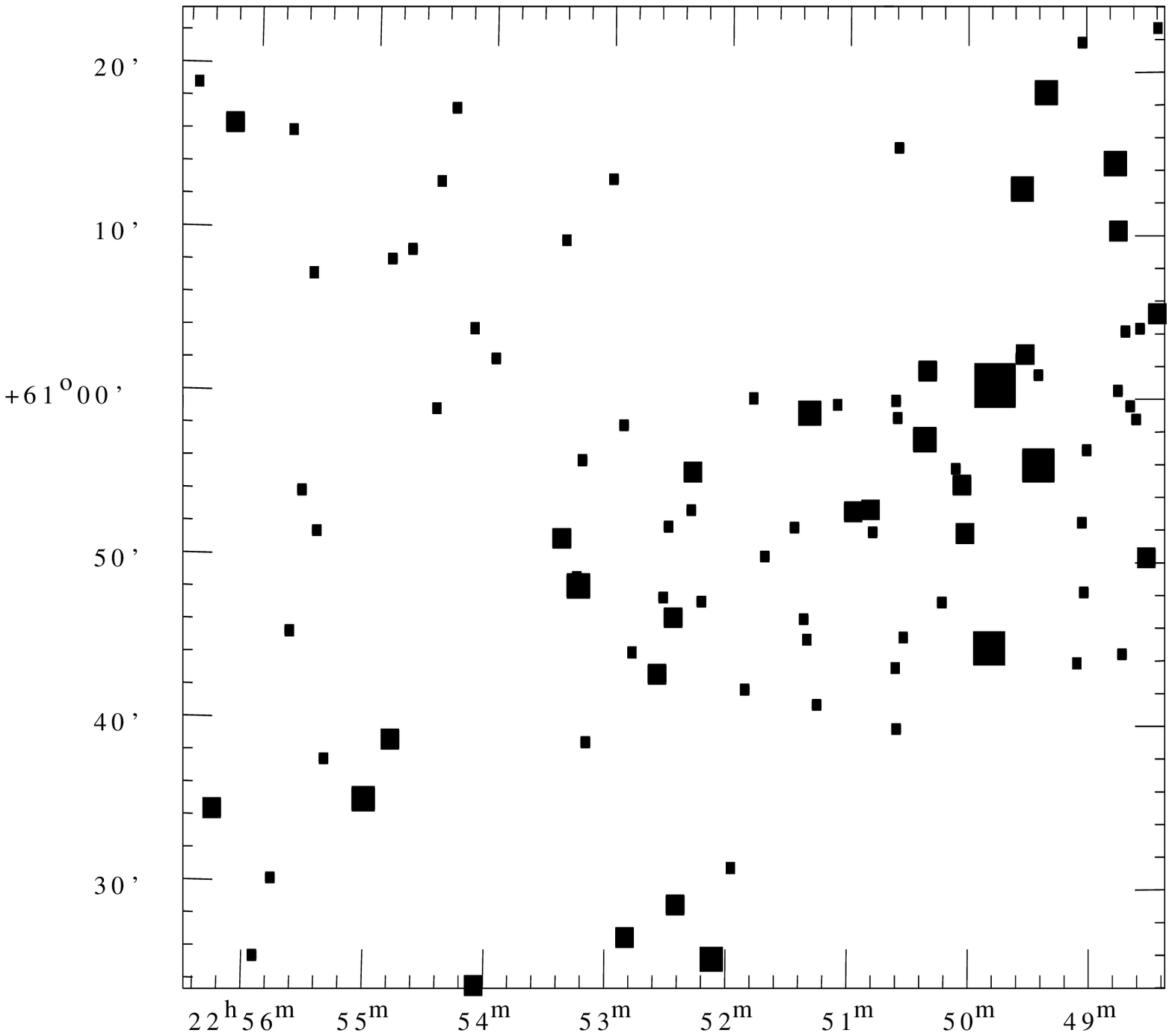} \hspace{4mm}
  \includegraphics[width=8.0cm]{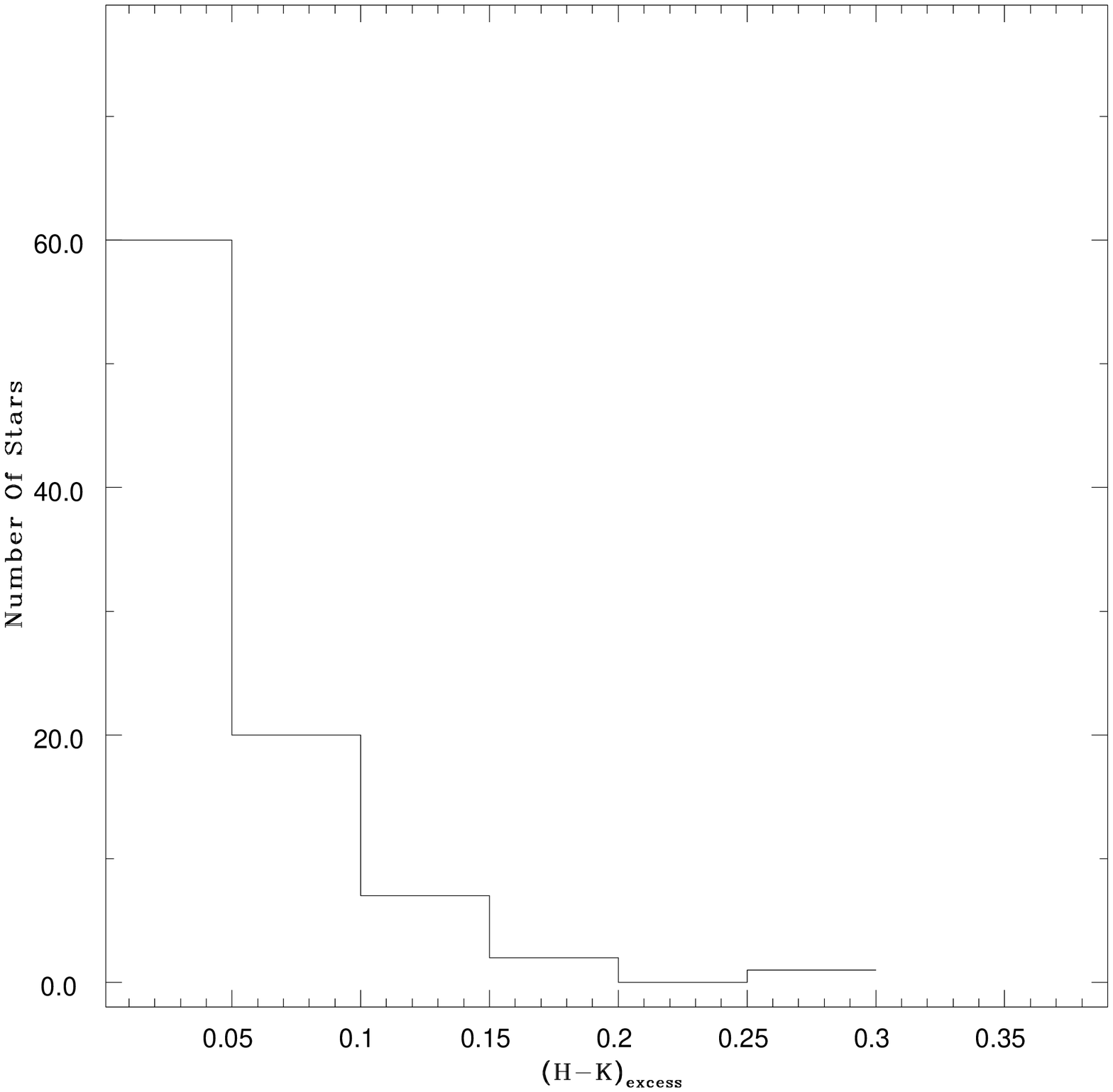}
  \caption{Age spread of YSOs.
   {\it left panel :} Spatial distribution of the $(H-K)_{\rm excess}$ of 
   the YSO candidates. The increasing sizes of filled boxes 
   represent the YSOs having $(H-K)_{\rm excess}$ in ranges 
   $0-0.05$ mag, $0.05-0.10$ mag, $0.10-0.15$ mag, $0.15-0.20$ mag 
   and $0.25-0.30$ mag, respectively.
  X-axis and Y-axis denote $\rm{RA_{J2000}}$ and $\rm{DEC_{J2000}}$, respectively. 
   {\it right panel:} Frequency distribution of the $(H-K)_{\rm excess}$ 
   for YSO candidates.}
  \label{fig:spaHk}
  \end{figure*}

The effect of the youthfulness of the YSOs in the spatial distribution 
can also be investigated with the help $(H-K)_{\rm excess}$, which 
is expected due to the presence of the circumstellar disk dissipating with 
time \citet{oasa06}. For YSOs, it is determined by the method used in
\citet{matsuyanangi06} and it lies in the range $0.002-0.3$ mag. 
The distribution of $(H-K)_{\rm excess}$ (Figure~\ref{fig:spaHk}, left panel)
does not show any spatial gradient with distribution of the YSOs,
indicating that the YSOs having different age are distributed 
uniformly within the cloud. Frequency distribution of the $H-K$ excess
is shown in Figure~\ref{fig:spaHk}, right panel and it indicates that 89\% 
of YSOs have nearly similar age. This represents a uniform distribution 
of the star formation rate within the cloud.

The morphology of the cloud is almost the same for all the YSOs 
and we find very little age dispersion in the YSOs.
There is no significant contribution of 
the external agents found in the formation of these YSOs in this cloud. 
Therefore, such a uniform distribution of YSO candidates might be the 
result of primordial fragmentation.

\section{Mass function}\label{sec:masFun}

\begin{table*}
\caption{The MF of the cluster NGC 7419. MS represent the MF for main-sequence population while PMS represents the MF for pre-main sequence population. The numbers of probable cluster members (N) have been
 obtained after subtracting the expected contribution of field stars in \S\ref{sec:colMag}. log$\phi$ 
represents log N(log m).}             
\label{tab:mf}      
\centering                          
\begin{tabular}{c c c c c c c c c}       
\hline                 
Magnitude range     & Mass  Range  & Mean &  \multicolumn{2}{c} { Inner region} &  \multicolumn{2}{c} { Outer region} &  \multicolumn{2}{c} { Whole region}\\
    ($V$ mag)      &   (M$_\odot$)  &  log (M$_\odot$)&  N & log$\phi$  &  N & log$\phi$  &  N & log$\phi$  \\
\hline
MS&&&&&&&&\\
\hline
15.0 - 16.0 &  8.59 - 7.10 &    0.894 &    9 & 2.036 &    6& 1.860 & 15&  2.258\\
16.0 - 17.0 &  7.10 - 5.20 &    0.789 &   11 & 1.912 &   16& 2.075 & 27&  2.302\\
17.0 - 18.0 &  5.20 - 3.57 &    0.642 &   14 & 1.933 &   27& 2.218 & 41&  2.400\\
18.0 - 19.0 &  3.57 - 2.45 &    0.479 &   16 & 1.991 &   34& 2.318 & 50&  2.486\\
19.0 - 20.0 &  2.45 - 1.78 &    0.326 &   13 & 1.972 &   65& 2.671 & 78&  2.750\\
20.0 - 21.0 &  1.78 - 1.44 &    0.207 &   14 & 2.172 &   96& 3.008 &110&  3.067\\
\hline
PMS&&&&&&&&\\
\hline
 & 7.00 - 6.00 & 0.544 & 4 & 1.505 & 7 & 1.748& 11& 1.945\\
 & 6.00 - 5.00 & 0.653 & 3 & 1.491 & 11& 2.055& 14& 2.160\\
 & 5.00 - 4.00 & 0.740 & 5 & 1.800 & 4 & 1.703&  9& 2.056\\
 & 4.00 - 3.00 & 0.067 & 5 & 1.873 & 16& 2.378& 21& 2.496\\
\hline
\end{tabular}
\end{table*}

\begin{figure}
  \includegraphics[width=8cm]{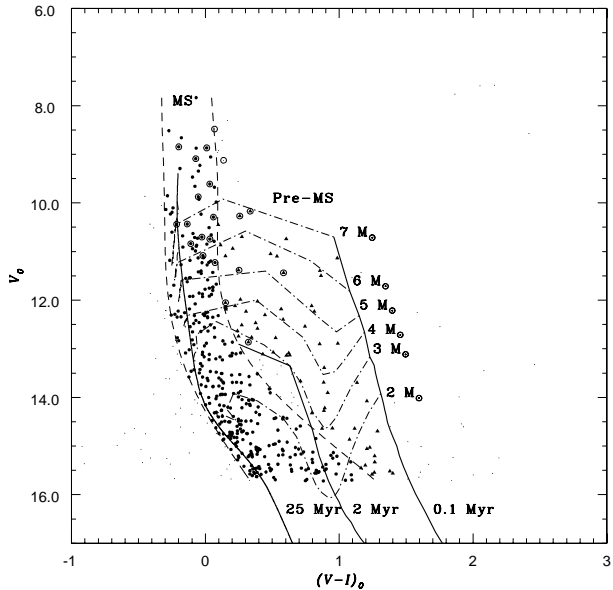}
  \caption{Extinction corrected CMD at mean $E(B-V)$=1.70 mag. Solid lines 
    represent the theoretical isochrones for main-sequence 
    at 25 Myr (GRD02) and pre-main-sequence at 2 Myr 
    and 0.1 Myr (SES00). Solid dots within 
    long-dash line show the stars selected for the estimation of 
    main-sequence MF at 25 Myr i.e. first episode of star formation, while 
    the triangles represent the stars used for pre-main-sequence MF. 
    Herbig Ae/Be are shown by open circles. Herbig Ae/Be stars within 
    main-sequence band (solid dots with open circles) have not been 
    considered in estimation of main-sequence MF and included in the 
    estimation of pre-main-sequence MF.}
  \label{fig:mfselect}
  \end{figure}

The initial mass function (IMF) is the distribution of stellar masses that form in a 
star formation event in a given volume of space. It is an  important result of star
formation and together 
with the star formation rate, the IMF dictates the evolution and fate of 
galaxies and star clusters \citep{kroupa02}.
However, the direct determination of IMF is not possible due to
the dynamical evolution of stellar systems. Therefore, we derived MF which is
 often expressed by the power law, $N(log m) \propto m^{\Gamma}$ and the slope of the MF is given 
as :

\begin{equation}
\Gamma =  d log N (log m) / d log m
\end{equation}

\noindent
where $N(log m)$ is the number of star per unit logarithmic
mass interval. In the solar neighborhood the classical value derived by 
\citet{salpeter55} is $\Gamma=-1.35$. In order to estimate the MF for the 
cluster region, we have used statistically cleaned sample described 
in \S\ref{sec:colMag}. We have divided statistically cleaned 
sample for two star formation episodes. 
The first star formation episode 
is at 25 Myr and the masses corresponding to $V\leq21$ mag have reached to the 
main-sequence, therefore  
only main-sequence MF have been considered for this episode.
In Figure~\ref{fig:mfselect}, the MS represents the 
main-sequence band (selected in \S\ref{sec:colMag}) and is drawn by long-dashed lines. 
The stars within this 
band are shown by solid dots. The main-sequence theoretical isochrone from 
GRD02 is shown by solid line at 25 Myr age with upper 
limit at $V\approx15.0$ mag because the stars above this point have been 
evolved from the main-sequence. Due to the above mentioned reasons, we have 
considered the stars with $V$ magnitude ranges $15-21$ mag for the 
main-sequence MF. The second episode is at age $\leq2$ Myr, therefore 
pre-main-sequence MF has been derived for the stellar population attached 
with this episode. The pre-main-sequence theoretical isochrones 
by SES00 have been drawn by solid lines for ages 2 Myr 
and 0.1 Myr (upper limit of age in model), and by dashed lines with dots 
for masses ranges from  2 $M\subsun$ to 7 $M\subsun$, respectively.
The stars which are lying in between this age range have been used for the 
estimate of pre-main-sequence MF and shown by triangles 
in Figure~\ref{fig:mfselect}. The Herbig Ae/Be stars, described 
in \S\ref{sec:nirDat}, are shown by open circles.  
All these stars are excluded in the estimation of main-sequence 
MF even though they are lying inside the selected MS band, 
and considered as pre-main-sequence stars and included in the estimation of pre-main-sequence MF. 
We have used the stars shown by triangles 
in Figure~\ref{fig:mfselect} and the Herbig Ae/Be stars inside 
the MS band for the estimation of pre-main sequence MF for 
masses down to $\sim3M\subsun$ due to the incompleteness of mass below this limit.  
 
\begin{figure}
  \includegraphics[height=4cm,width=6cm,angle=270]{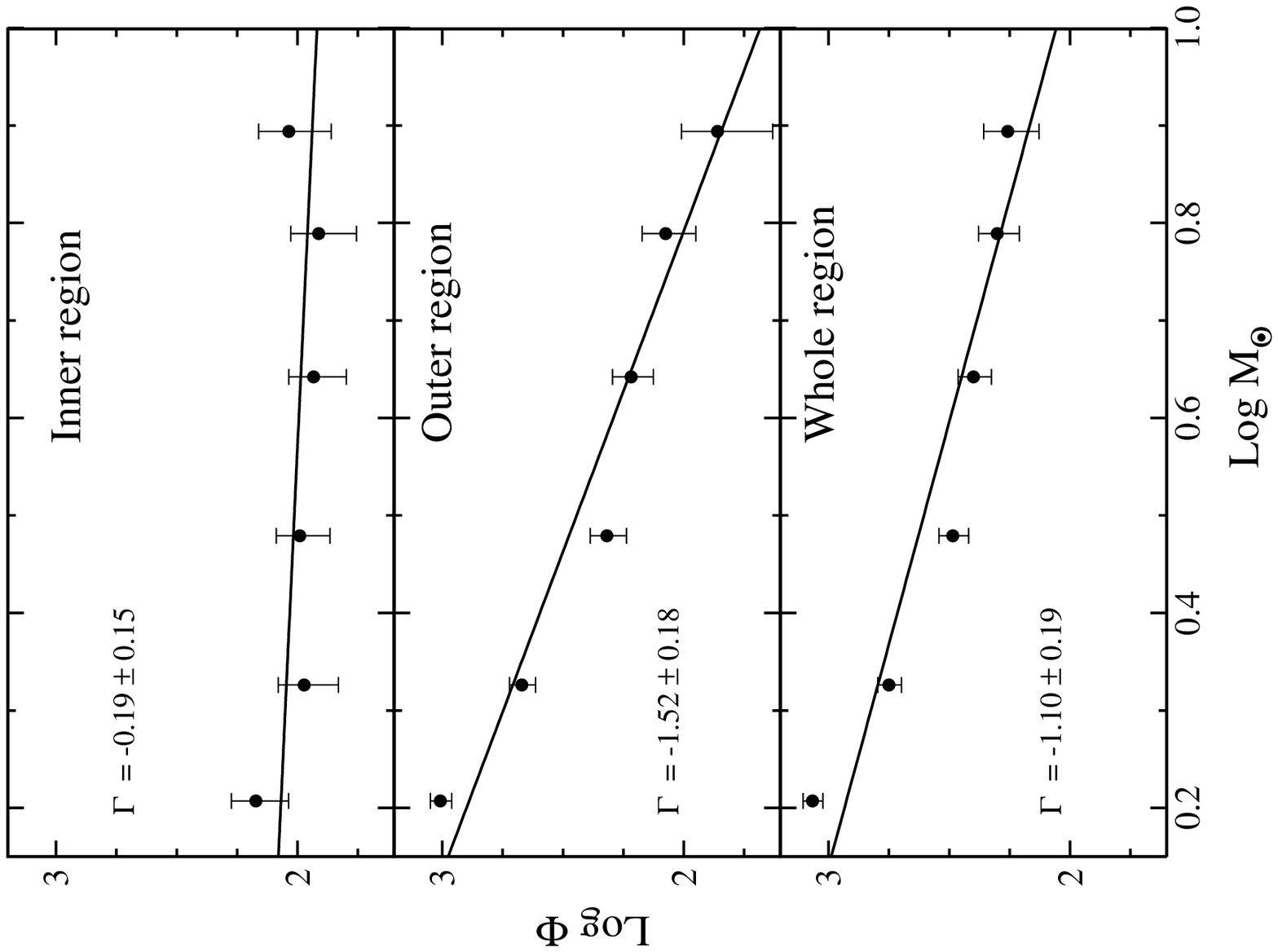} 
  \includegraphics[height=4cm,width=6cm,angle=270]{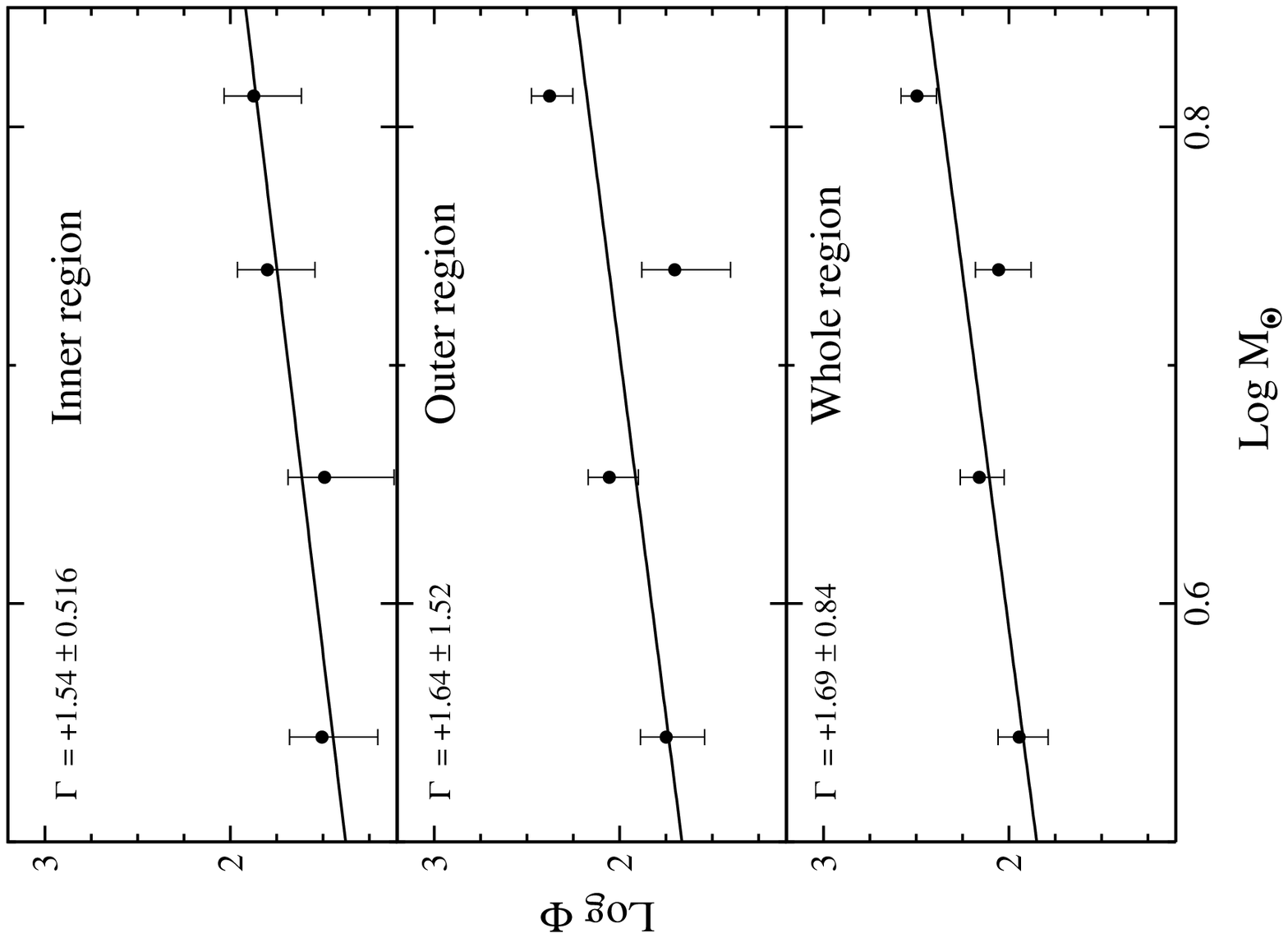}
  \caption{MF determination of MF in the various subregions of the 
    cluster NGC 7419. $left panel$ : for main-sequence and, 
    for pre-main-sequence.
    $right panel$ : Inner region represents the core region ($r\leq 1\arcmin$ ), Outer 
    region represents the corona 
    region ($  1\farcm0 \leq r \leq 3\farcm5$ ) and Whole region 
    represents the cluster region ($ r \leq 3\farcm5$), as defined 
    using radial density profile. The error bars represent $1/\sqrt N$ 
    errors.} 
  \label{fig:mf}
  \end{figure}

We have derived the MF from LF using the theoretical evolutionary models.
The MF in two subregions i.e., the inner ($r\leq1\arcmin$) and the 
outer (1\arcmin$ \leq r \leq $3\farcm5) region of the cluster, 
respectively, as well as for the whole cluster region 
($r\leq3\farcm5$) are given in Table~\ref{tab:mf} and plotted 
in Figure~\ref{fig:mf}.

The slope of the MF of the main-sequence stars in the inner region 
of the cluster is estimated as $\Gamma=-0.19\pm0.15$, which
is flatter than the \citet{salpeter55} value, while for the stars in the outer 
region, $\Gamma=-1.52\pm0.18$, which is  similar to the \citet{salpeter55} 
value. The $\Gamma$ value for the stars is thus steeper in the outer region 
indicating a preferential distribution of relatively massive stars towards 
the cluster center indicating a segregation of stellar masses. The slope for 
the MF for the whole cluster is $\Gamma=-1.10\pm0.19$, which is similar 
to the Salpeter value in the mass range of $8.6-1.4 M\subsun$. 
This value of the MF is similar to the value estimated 
by \citet{beauchamp94} i.e. $\Gamma=-1.25\pm0.10 $ estimated at 
age$\sim$14 Myr and distance $\sim2.3$ kpc for the cluster region.

The slopes of the pre-main-sequence MF for the intermediate mass 
range $7-3M\subsun$ in the cluster are derived 
as  $\Gamma=+1.54\pm0.52$ for inner, $\Gamma=+1.64\pm1.52$ for 
outer, and $\Gamma=+1.68\pm0.84$ for whole cluster region. The value of 
the pre-main-sequence MF  slope is significantly flatter than the Salpeter value. 


\section{Mass segregation}\label{sec:masSeg}
 
There is evidence for mass segregation in a few Galactic as well as LMC 
star clusters, with the highest mass stars preferentially found towards 
the center of the 
cluster \citep[see][]{sagar88,sagar91,sagar98,pandey01,pandey05,kumar08}. 
It is a well accepted finding that stars in 
the clusters evolve rapidly towards a state of energy equipartition through 
stellar encounters i.e., mass segregation. However, observations of very 
young clusters \citep[e.g.,][]{sagar88,pandey92,pandey05,hillenbrand97} suggest 
that the mass segregation may be the imprint of star formation itself.

\begin{figure}
  \includegraphics[width=8cm]{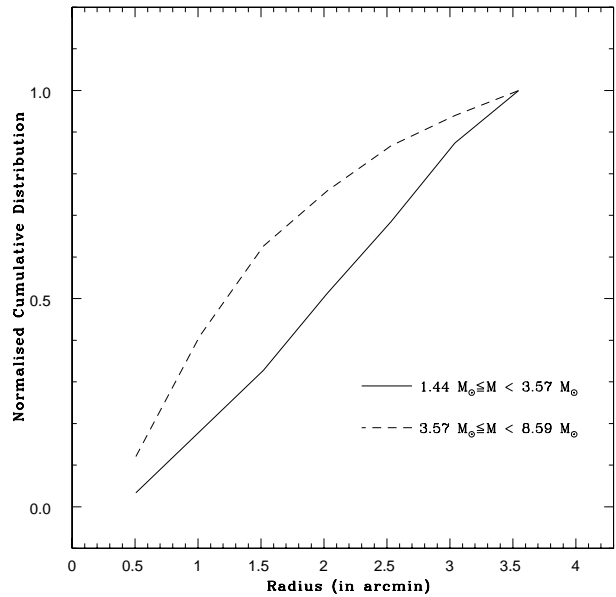}
  \caption{Cumulative radial distribution of main-sequence stars in mass 
    intervals $1.44 \leq M/M\subsun < 3.57$ 
    and  $3.57 \leq M/M\subsun < 8.59$.} 
  \label{fig:mseg}
  \end{figure}

We have subdivided the main-sequence sample (used for the main-sequence MF 
in \S\ref{sec:masFun}) into two mass groups 
($3.6 \leq M/M\subsun < 8.6$;
$1.4 \leq M/M\subsun < 3.6$) to characterize the degree of 
mass segregation in NGC 7419. Figure~\ref{fig:mseg} shows cumulative 
distribution of main-sequence stars as a function of radius in two different 
mass groups. Effect of mass segregation can be seen in this figure
in the sense that more massive stars 
($3.6 \leq M/M\subsun < 8.6$) tend to lie towards the 
cluster center. The Kolmogorov-Smirnov (KS) test confirms the statement 
that the cumulative distribution of massive stars in the cluster is different 
from that of relatively less massive stars at a 
confidence level of 99\%. There is a strong evidence of mass segregation 
in the main-sequence stars within this cluster. We estimated the relaxation 
time to decide whether the mass segregation discussed above is primordial or 
due to dynamical relaxation. To estimate the dynamical relaxation time $T_E$, 
we have used the relation

\begin{equation}
T_E = \frac {8.9 \times 10^5 N^{1/2} {R_h}^{3/2}} { {\bar{m}}^{1/2} log(0.4N) }
\end {equation}

\noindent
where $N$ is the number of cluster stars, $R_h$ is the radius containing half
of the cluster mass and $\bar {m}$ is the average mass of the cluster stars 
\citep{spitzer71}. The total number of stars in the cluster region 
are estimated as 321 in the mass 
range ($1.4 \leq M/M\subsun < 8.6$). 
For the half mass radius, we have used half of the cluster 
extent i.e., 1.63 pc. Taking average mass of the cluster 
as $3.92~M\subsun$, we have estimated
the dynamical relaxation time $T_E$ for the cluster as 8.0 Myr, which
is lower than the turn-off age of the cluster i.e., 19 -- 25 Myr.
Therefore, we can conclude that the cluster is dynamically relaxed.

\section{X-ray Data Analysis}\label{sec:xraDat}

\begin{table}
\caption{XMM-Newton Observations of NGC 7419.}             
\label{tab:xmmlog}      
\begin{tabular}{c c }        

\hline                 
Parameter              & NGC 7419                               \\
\hline
Observation ID         & 0201160501                              \\
Start time (UT)        & 02 Feb 2004  02:40:39                    \\
Stop time  (UT)        & 02 Feb 2004  14:19:16                    \\
Usable time(ks)        & 39.62(MOS1),40.21(MOS2),35.35(PN)   \\
EPIC mode              & full frame                               \\
Optical filter         & Medium                                   \\
\hline
\end{tabular}
\end{table}

We have used the archival X-ray data from the XMM-Newton observations 
of NGC 7419. The observations were proposed by Christian Motch to search 
for Be+white dwarfs binaries in NGC 7419. The observation details are 
given in Table~\ref{tab:xmmlog}. 
Our analysis is based on the CCD images from the European Photon 
Imaging Camera (EPIC). The details of the X-ray telescope and EPIC 
PN and MOS cameras are given by \citet{jansen01}, \citet{struder01} and 
\citet{turner01}. Data reduction followed standard procedures using 
the XMM-Newton Science Analysis System software (SAS; version 7.0.0). 
Event files for MOS and PN are generated by using tasks $\sc emchain$ 
and  $\sc epchain$, respectively. Data from 
the three cameras were individually screened for high background episodes 
and the time intervals during which the total count rate (for 
single events of energy above 10 keV) in the instruments
exceeded 0.35 and 1.0 counts s$^{-1}$ for the MOS and PN detectors, 
respectively, were excluded.

\subsection{Source detection and identification} \label{sec:sorDet}
  \begin{figure}
  \includegraphics[width=8cm]{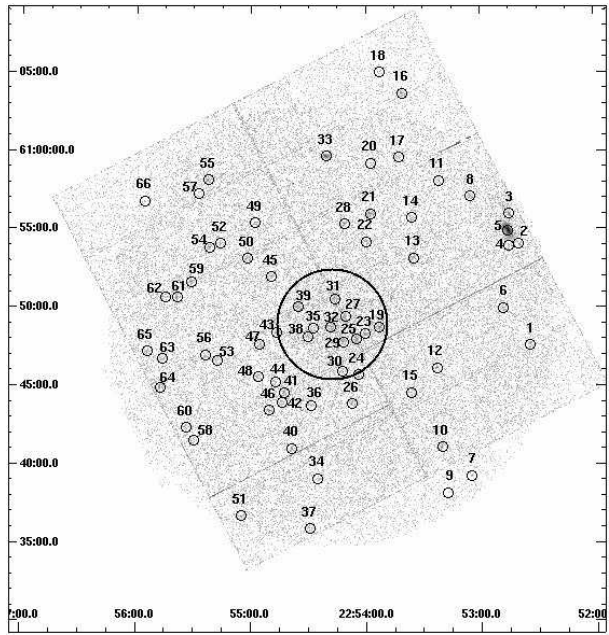}
  \caption{The X-ray  point sources in the mosaic image of MOS and 
    PN detector in the energy band 0.3 -- 7.5 keV with their ID number 
    as given in Table~\ref{tab:xraysrc}. 
    X-axis and Y-axis denote $\rm{RA_{J2000}}$ and $\rm{DEC_{J2000}}$, respectively. } 
  \label{fig:xrayimg}
  \end{figure}

\begin{table*}
\scriptsize
\caption{X-ray Sources detected using SAS task $edetect\_chain$. 
  Column 1 (XID), represents the identification number from X-ray source 
  detection and display in Figure~\ref{fig:xrayimg},
  Column 2 (Position), shows source position from X-ray source detection,  
  Column 3 (Count Rates), represents the pn-equivalent count rates in 
  energy band 0.3 -- 7.5 keV, 
  Column 4 (Hardness ratio), $HR1$ and $HR2$, defined in text, the 
  sign of $\leq$ represents the hardness ratio calculated 
  after replacing zero counts by upper limits in either of the energy band, 
  Column 5 (Remarks), for PCM (the X-ray sources with membership 
  probability using Hardness Ratio -- the X-ray sources inside the dotted
  box in Figure~\ref{fig:hratio}) IR (the X-ray sources with NIR counterparts 
  within 6\arcsec search radius using 2MASS data)
  , OPT (the X-ray sources with optical counterparts 
  within $6^{\prime\prime}$ search radius using our $UBVRI$ photometry.)
  , Cl (the X-ray sources within cluster region NGC 7419 
  -- see \S\ref{sec:radDen})}
\label{tab:xraysrc}      

\begin{tabular}{c c c c c c l } \hline
 XID   &$\rm{RA_{J2000}}$ & $\rm{DEC_{J2000}}$           &  Count Rates                 &  \multicolumn{2}{c}{Hardness ratio} &  Remarks\\ 
       &  (h  m  s)       & (\degr~ \arcmin~ \arcsec)    & $\rm{10^{-3}~counts~s^{-1}}$ & $HR1$ & $HR2$                       &\\
\hline
     1&	22 52 35&  60 47 37 &   10.92$\pm $ 1.50  &	-0.40$\pm $  0.170 &	0.34	$\pm$ 0.21 &  IR \\
     2&	22 52 40&  60 54 06 &    1.44$\pm $ 0.58  &	0.25 $\pm $  0.393 &	$\leq$-0.20	 &      ------\\
     3&	22 52 45&  60 56 01 &   17.97$\pm $ 1.58  &	0.16 $\pm $  0.098 &	-0.19	$\pm$ 0.11 &  IR \\
     4&	22 52 46&  60 54 00 &    2.61$\pm $ 0.74  &	-0.01$\pm $  0.276 &	-0.58	$\pm$ 0.48 &    ------\\
     5&	22 52 46&  60 54 57 &  294.66$\pm $ 5.33  &	0.47 $\pm $  0.023 &	-0.37	$\pm$ 0.02 &  IR \\
     6&	22 52 49&  60 50 01 &    3.89$\pm $ 0.73  &	0.82 $\pm $  0.357 &	0.28	$\pm$ 0.19 &  IR, PCM \\
     7&	22 53 05&  60 39 19 &    4.83$\pm $ 2.06  &	$\leq$-0.43        &	$\leq$0.64	   &    ------ \\
     8&	22 53 06&  60 57 08 &    4.15$\pm $ 0.83  &	0.19 $\pm $  0.512 &	0.77	$\pm$ 0.26 &    ------ \\
     9&	22 53 17&  60 38 13 &    2.82$\pm $ 1.82  &	0.54 $\pm $  0.565 &	$\leq$-0.63	   &  IR \\
    10&	22 53 21&  60 41 09 &    5.05$\pm $ 0.73  &	0.21 $\pm $  0.167 &	-0.10	$\pm$ 0.17 &  IR \\
    11&	22 53 22&  60 58 08 &    2.20$\pm $ 0.57  &	0.43 $\pm $  0.255 &	$\leq$-0.43	   &    ------ \\
    12&	22 53 23&  60 46 12 &    3.91$\pm $ 0.54  &	0.12 $\pm $  0.145 &	-0.34	$\pm$ 0.18 &  IR \\
    13&	22 53 35&  60 53 12 &    3.75$\pm $ 0.59  &	0.71 $\pm $  0.683 &	0.83	$\pm$ 0.21 &   PCM\\
    14&	22 53 36&  60 55 47 &    3.38$\pm $ 0.61  &	0.69 $\pm $  0.421 &	0.54	$\pm$ 0.21 &   PCM\\
    15&	22 53 37&  60 44 37 &    0.84$\pm $ 0.31  &	$\leq$0.64         &	0.67	$\pm$ 0.44 &   PCM\\
    16&	22 53 41&  61 03 43 &   18.95$\pm $ 2.03  &	0.40 $\pm $  0.118 &	-0.50	$\pm$ 0.14 &  IR \\
    17&	22 53 42&  60 59 38 &    3.18$\pm $ 0.75  &	0.04 $\pm $  0.233 &	-0.10	$\pm$ 0.31 &  IR \\
    18&	22 53 52&  61 05 04 &    6.12$\pm $ 1.14  &	$\leq$0.97         &	0.71	$\pm$ 0.23 &  PCM\\
    19&	22 53 53&  60 48 48 &    5.45$\pm $ 0.55  &	0.83 $\pm $  0.331 &	0.70	$\pm$ 0.12 &  IR, OPT, PCM, Cl\\
    20&	22 53 57&  60 59 14 &    1.32$\pm $ 0.51  &	0.58 $\pm $  0.406 &	-0.41	$\pm$ 0.46 &  IR \\
    21&	22 53 57&  60 56 02 &    8.79$\pm $ 0.96  &	$\leq$0.85         &	0.88	$\pm$ 0.15 &  IR, PCM\\
    22&	22 54 00&  60 54 15 &    1.56$\pm $ 0.41  &	$\leq$0.70         &	0.63	$\pm$ 0.31 &  PCM\\
    23&	22 54 00&  60 48 22 &    2.56$\pm $ 0.43  &	$\leq$0.70         &	0.75	$\pm$ 0.21 &  OPT, PCM, Cl\\
    24&	22 54 04&  60 45 45 &    0.69$\pm $ 0.24  &	0.08 $\pm $  0.321 &	0.13	$\pm$ 0.19 &  IR, OPT \\
    25&	22 54 05&  60 48 04 &    5.68$\pm $ 0.71  &	0.07 $\pm $  0.183 &	0.48	$\pm$ 0.16 &  IR, OPT, Cl \\
    26&	22 54 07&  60 43 54 &    1.75$\pm $ 0.37  &	0.28 $\pm $  0.230 &	-0.22	$\pm$ 0.26 &  IR, OPT \\
    27&	22 54 10&  60 49 29 &    7.18$\pm $ 1.17  &	-0.56$\pm $  0.352 &	0.84	$\pm$ 0.25 &  Cl\\
    28&	22 54 11&  60 55 23 &    5.96$\pm $ 0.78  &	0.30 $\pm $  0.137 &	-0.51	$\pm$ 0.18 &  IR \\
    29&	22 54 12&  60 47 51 &    1.64$\pm $ 0.37  &        $\leq$-0.48     &	0.75	$\pm$ 0.28 &  IR, OPT, Cl \\
    30&	22 54 12&  60 45 59 &    1.62$\pm $ 0.33  &	0.09 $\pm $  0.20 &	-0.98	$\pm$ 0.39 &  IR, OPT, Cl \\
    31&	22 54 16&  60 50 36 &    5.61$\pm $ 0.73  &	-0.16$\pm $  0.17 &	0.41	$\pm$ 0.17 &  IR, OPT, Cl \\  
    32&	22 54 18&  60 48 48 &    4.78$\pm $ 0.71  &        $\leq$0.37      &	0.53	$\pm$ 0.17 &  IR, OPT, Cl \\
    33&	22 54 21&  60 59 42 &   24.48$\pm $ 1.52  &	0.02 $\pm $  0.06 &	-0.65	$\pm$ 0.10 &   ------\\
    34&	22 54 25&  60 39 07 &    1.66$\pm $ 0.40  &	0.49 $\pm $  0.25 &	$\leq$-0.25	   &  IR \\
    35&	22 54 27&  60 48 44 &    1.40$\pm $ 0.38  &	0.60 $\pm $  0.52 &	0.50	$\pm$ 0.31 &  PCM, Cl \\
    36&	22 54 29&  60 43 49 &    1.64$\pm $ 0.55  &	-0.06$\pm $  0.29 &	$\leq$-0.18	   &  IR, OPT \\
    37&	22 54 29&  60 35 57 &    7.26$\pm $ 2.37  &	$\leq$-0.24       &	$\leq$0.84	   &   ------\\
    38&	22 54 30&  60 48 12 &    1.54$\pm $ 0.36  &	0.85 $\pm $  0.69 &	0.65	$\pm$ 0.28 &  OPT, PCM, Cl\\
    39&	22 54 35&  60 50 05 &    5.96$\pm $ 1.05  &	0.60 $\pm $  0.20 &	-0.43	$\pm$ 0.21 &  OPT, Cl \\
    40&	22 54 39&  60 41 03 &    4.48$\pm $ 0.59  &	0.32 $\pm $  0.14 &	-0.58	$\pm$ 0.19 &  IR \\
    41&	22 54 42&  60 44 38 &    3.33$\pm $ 0.84  &	$\leq$-0.14        &	$\leq$0.79 &  IR, OPT  \\
    42&	22 54 44&  60 44 01 &    1.50$\pm $ 0.39  &	$\leq$0.46        &	$\leq$0.60	 &    ------\\
    43&	22 54 46&  60 48 27 &    1.09$\pm $ 0.32  &	0.37 $\pm $  0.91 &	0.80	$\pm$ 0.38 &  ------ \\
    44&	22 54 47&  60 45 18 &    1.25$\pm $ 0.34  &	$\leq$0.78        &	$\leq$0.56	 &   PCM\\
    45&	22 54 49&  60 52 01 &    1.14$\pm $ 0.35  &	$\leq$0.78        &	$\leq$0.53	&    PCM\\
    46&	22 54 50&  60 43 29 &    3.78$\pm $ 0.54  &	0.93 $\pm $  0.83 &	0.87	$\pm$ 0.19 &   OPT, PCM\\
    47&	22 54 55&  60 47 42 &    2.25$\pm $ 0.42  &	0.31 $\pm $  0.21 &	-0.03	$\pm$ 0.22 &   IR, OPT  \\
    48&	22 54 56&  60 45 37 &    1.22$\pm $ 0.36  &	0.43 $\pm $  0.40 &	0.23	$\pm$ 0.32 &   IR, OPT \\
    49&	22 54 58&  60 55 27 &    2.91$\pm $ 0.56  &	$\leq$0.74        &	0.88	$\pm$ 0.29 &   PCM\\
    50&	22 55 02&  60 53 11 &    1.40$\pm $ 0.48  &	$\leq$0.86        &	$\leq$0.60	 &   PCM\\
    51&	22 55 05&  60 36 45 &    4.92$\pm $ 2.32  &	$\leq$0.93        &	$\leq$0.79	 &   PCM\\
    52&	22 55 16&  60 54 07 &    3.07$\pm $ 0.60  &	0.97 $\pm $  0.72 &	0.72	$\pm$ 0.24 &   IR, OPT, PCM\\
    53&	22 55 17&  60 46 39 &    0.92$\pm $ 0.37  &	$\leq$0.70        &	0.82	$\pm$ 0.52 &   IR, OPT, PCM\\
    54&	22 55 22&  60 53 52 &    3.12$\pm $ 0.61  &	$\leq$0.96        &	0.63	$\pm$ 0.23 &   PCM\\
    55&	22 55 22&  60 58 10 &   12.19$\pm $ 1.22  &	0.01 $\pm $  0.11 &	-0.30	$\pm$ 0.14 &   IR \\
    56&	22 55 24&  60 47 02 &    3.35$\pm $ 0.97  &	-0.23$\pm $  0.64 &	0.76	$\pm$ 0.39 &   ------\\
    57&	22 55 27&  60 57 18 &    2.14$\pm $ 0.61  &	0.20$\pm $  0.27  &	-0.70	$\pm$ 0.44 &   ------\\
    58&	22 55 30&  60 41 35 &    4.14$\pm $ 1.05  &	-0.62$\pm $  0.41 &	0.74	$\pm$ 0.42 &   ------\\
    59&	22 55 31&  60 51 39 &    6.48$\pm $ 1.25  &	-0.46$\pm $  0.45 &	0.84	$\pm$ 0.28 &   ------\\
    60&	22 55 34&  60 42 23 &    1.85$\pm $ 0.55  &	$\leq$0.99        &	$\leq$0.54	 &    PCM\\
    61&	22 55 38&  60 50 42 &    4.95$\pm $ 0.77  &	$\leq$0.87        &	$\leq$0.79	 &    PCM\\
    62&	22 55 44&  60 50 43 &    5.48$\pm $ 1.08  &	-0.76$\pm $  0.44 &	0.91	$\pm$ 0.32 &   ------\\
    63&	22 55 46&  60 46 48 &    7.41$\pm $ 0.48  &	$\leq$0.93        &	$\leq$-0.12	 &    PCM\\
    64&	22 55 47&  60 44 58 &    9.86$\pm $ 1.08  &	0.58 $\pm $  0.13 &	-0.42	$\pm$ 0.13 &   IR \\
    65&	22 55 54&  60 47 14 &    2.22$\pm $ 0.63  &	0.10 $\pm $  1.50 &	0.94	$\pm$ 0.39 &   ------\\
    66&	22 55 56&  60 56 50 &    2.94$\pm $ 0.83  &	-0.02$\pm $  0.30 &	0.14	$\pm$ 0.37 &   IR \\
\hline                                   
\end{tabular}
\end{table*}

The source detection is based on the SAS point source detection 
algorithm  $\sc edetect\_chain$. Three energy ranges were selected -- a 
soft ($S_X$) band (0.3 -- 0.7 keV), a medium ($M_X$) band (0.7 -- 1.2 keV) 
and a hard ($H_X$) band (1.2 -- 7.5 keV), and we built the corresponding 
images for the different instruments of the EPIC. Finally, source detection 
was performed on these images using $\sc edetect\_chain$ task, which is a chain 
script of various sub tasks.

We inspected each source manually to reject false detections due to the 
instrumental artifacts, from the final list created by $\sc emldetect$ task.
In this way, we found 66 sources with a combined maximum likelihood value
greater than 10 in all three instruments.  The details of all 
the genuine X-ray sources in the energy band 0.3 -- 7.5 keV are tabulated 
in Table~\ref{tab:xraysrc} and shown in Figure~\ref{fig:xrayimg}. 
 
We cross-correlated the
X-ray source list with our optical $UBVRI$ photometric source list
and  NIR 2MASS All Sky data release data \citep{cutri03}  
within the search radius 6\arcsec.
The position of the stars in optical CCD are converted
into $\rm RA_{J2000}$ and $\rm DEC_{J2000}$
using the Guide Star Catalogue- II (GSC 2.2, 2001). 
There are 18 X-ray sources with optical counterparts
and 31 with NIR counterparts found after cross-correlation.
Furthermore, none of the Herbig Ae/Be
stars is having an X-ray counterpart.
The unidentified X-ray sources i.e., without having any optical or NIR 
counterparts within XMM-Newton field of view 
($30\arcmin \times 30\arcmin$), are 31 in number (47\% of the total).

\subsection{The detection limit}\label{sec:detLimit}

  \begin{figure*}
  \includegraphics[width=8cm]{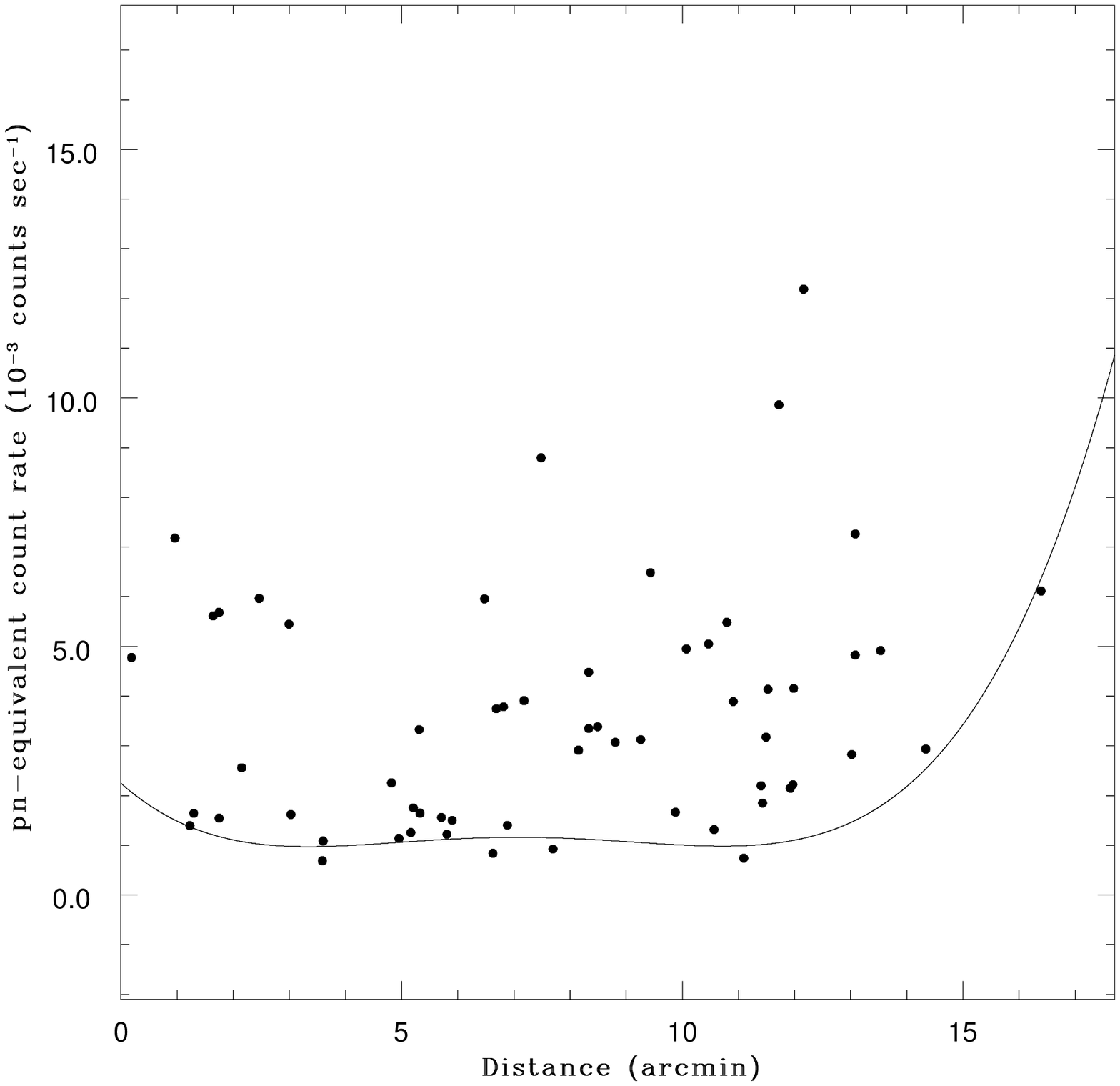} \hspace{4mm}
  \includegraphics[width=8cm]{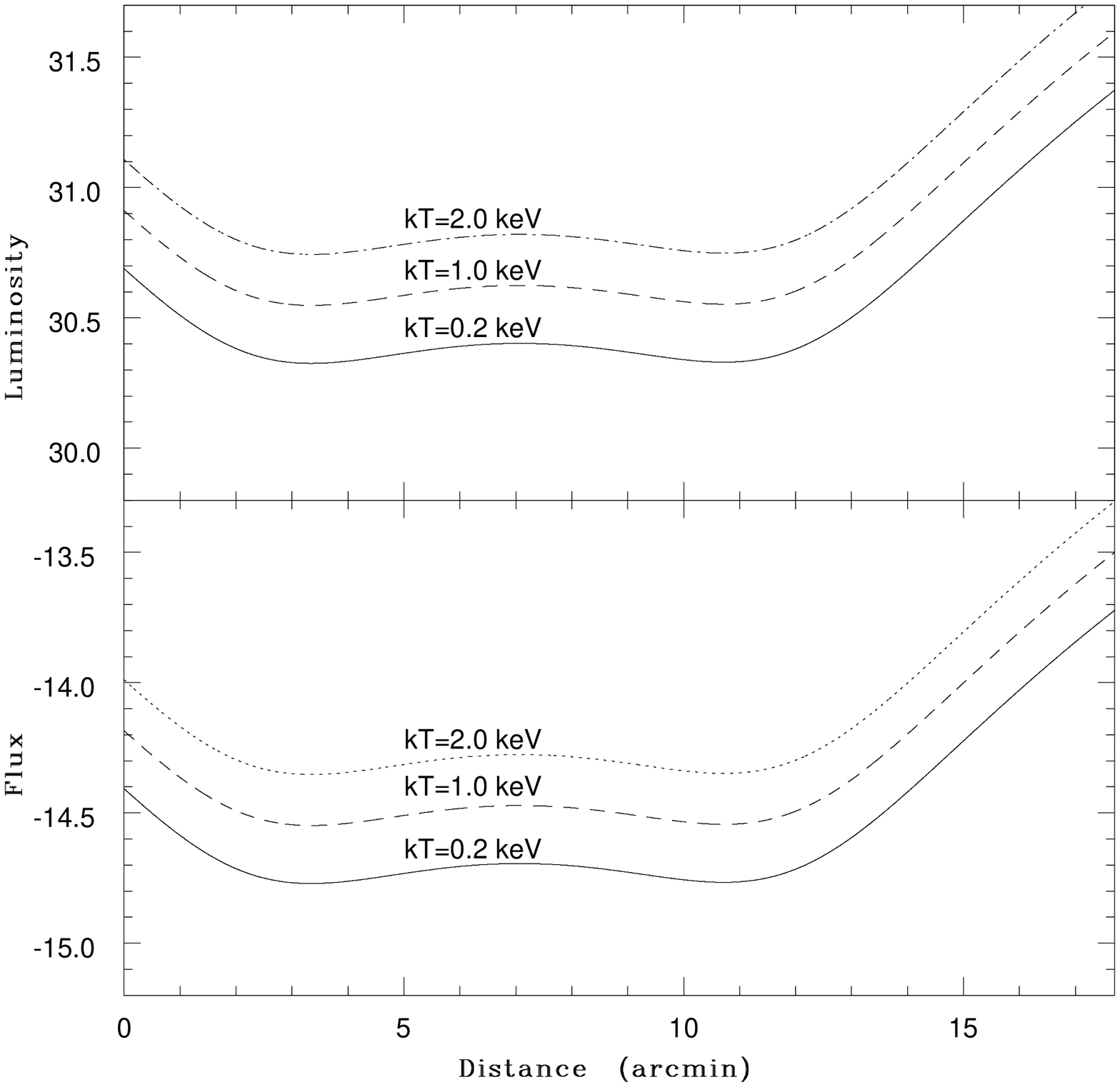}
  \caption{The variation of X-ray detection limits within XMM-Newton field of 
    view in energy band 0.3 --- 7.5 keV. $left panel$ : The distribution of the 
    pn-equivalent count rates of the X-ray sources as a function of their 
    distance from the center of the cluster NGC 7419.
    The solid line shows the adopted lower limit given in equation (4).
    (b)$right panel$:  $Lower~panel$ : estimated detection limit expressed in terms of the 
    observed flux (log(erg s$^{-1}$ cm$^{-2}$)), $Upper~panel$ : the 
    detection limit, expressed in luminosity (log(erg s$^{-1}$))
    using the distance of the cluster NGC 7419.}
  \label{fig:detLimit}
  \end{figure*}

The analysis of detection limits in the XMM-Newton field of view is equally 
important to decide the X-ray emission level of undetected Herbig Ae/Be stars,
as none of the Herbig Ae/Be stars is emitting X-rays. The faintest source 
detected above $2\sigma$ in the XMM-Newton field of view (FOV) has a 
count rate of $8.4\times 10^{-4}$ counts s$^{-1}$ in the PN detector. 
The detection limit is not constant throughout the field of view because 
of the following three reasons: (a) a gap between the CCDs, and  unoverlapped 
detector areas leading to a non-uniform effective detection area, (b) the 
decreasing exposure duration from the FOV center to its edges, causing 
a non-uniformity in the effective exposure times as well, and (c) the dense 
population of the stars near the center of the cluster affecting the 
detection limit non-uniformly. We have neglected the effects of the gap 
between CCDs, as an approximation, and investigated the effects of other 
parameters on the detection limit.

We analyzed the exposure map created by the $\sc expmap$ task for the PN detector 
in the energy band 0.3 -- 7.5 keV to derive the effect of exposure duration.
It displays a smooth decrease in the counts from the center of the FOV to
its edges by about a factor of three. 
 If we are not considering 
the background effect, the signal-to-noise ratio 
is  smaller by a factor of $\sqrt{3}\approx1.7$
for the source at the edges of the FOV 
compare to the source at the center.
Therefore, the detection limit in the edges of the FOV will be twice  the 
detection limit at the center of the FOV due to the non-uniformity of exposure 
duration.  

The variations in the detection limit due to the
dense clustering at the center of FOV have been determined 
using the  approach by \citet{sana06}.
We computed an equivalent PN count rate in the energy band 0.3 -- 7.5 keV
due to the gap between CCDs. An empirical relation between count rates in 
the PN and MOS detectors was calculated and  found to be approximately linear.
For the sources which fall in the gaps between the PN CCDs, this relation is 
used  to convert the MOS count rates into the PN count rates and
these PN count rates are called pn-equivalent count rates. 
Figure~\ref{fig:detLimit}, left panel, displays the source pn-equivalent 
count rates as a function of the distance from the center of the FOV i.e., 
the center of the cluster NGC 7419. We adjusted a four-degree polynomial 
by selecting the faintest sources to derive an empirical detection limit 
in terms of pn-equivalent count rates ($cr_{lim}$) as a function of 
distance ($d$) from the center of the FOV. 
This detection limit (in units of $\rm{10^{-3}~counts~s^{-1}}$)
is shown by a solid line in Figure~\ref{fig:detLimit}, left panel and 
described by the following relation:

  \begin{equation}
     cr_{lim}(d)= 2.246 - 0.999 d + 
      0.267 d^{2} -0.028 d^{3} +0.001 d^{4}
     \label{eqn:det}
  \end{equation}

\noindent
where the distance $d$ is expressed in arcmin from the center of the FOV.
Left panel of Figure~\ref{fig:detLimit} shows that the detection limit is 
higher in center of the FOV ($d<3\farcm5$), which is the cluster region as 
decided in \S\ref{sec:radDen}. It might be the effect of the extended X-ray 
emission from the cluster region, discovered in \S\ref{sec:extXra}.
The detection limit increases towards the CCD edges, as estimated by the 
exposure map. Using mission count rate simulator, 
WebPIMMS\footnote{WebPIMMS is a NASA's HEASARC tool powered by PIMMS 
version 3.9. It is hosted 
at URL: http://heasarc.gsfc.nasa.gov/Tools/w3pimms.html}, we converted 
the pn-equivalent count rates into fluxes and luminosities.
Assuming a hydrogen column density ($N_H$) of $\rm{9.9\times10^{21}~cm^{-2}}$,
at $E(B-V)=$1.70 mag for the cluster NGC 7419,
the Raymond-Smith single temperature optically thin thermal plasma models 
were used  at temperatures (kT) of 0.2 keV, 1.0 keV and 2.0 keV.
Results are displayed in Figure~\ref{fig:detLimit}, right panel for different 
plasma temperatures. Luminosities are calculated for the distance of the 
cluster NGC 7419 i.e., 3.2 kpc. Within the cluster region ($d<3\farcm5$), 
the detection limit of flux varies from 
$1.66\times10^{-15}$ to $3.89\times10^{-15}$ $\rm{erg~cm^{-2}~s^{-1}}$for  
soft X-ray sources at kT=0.2 keV, 
$3.02\times10^{-15}$ to $6.31\times10^{-15}$ $\rm{erg~cm^{-2}~s^{-1}}$ for   
X-ray sources at kT=1.0 keV, and
$4.47\times10^{-15}$ to $10.0\times10^{-15}$ $\rm{erg~cm^{-2}~s^{-1}}$ for   
X-ray sources at kT=2.0 keV. We have estimated the detection limit of 
21 Herbig Ae/Be stars, according to their positions in 
the cluster region using equation~\ref{eqn:det}.  As majority of 
Herbig Ae/Be stars have X-ray temperatures greater than 1.0 keV 
\citep{stelzer06, hamaguchi05}, the detection limits for these stars are 
estimated as the median of the limiting X-ray luminosity ($\rm {L_X}$) of Herbig Ae/Be stars  
i.e., $5.20\times10^{30}$ $\rm{erg~s^{-1}}$.

\subsection{Probable Members Using Hardness Ratio}\label{sec:proMem}

  \begin{figure*}
  \includegraphics[width=8cm]{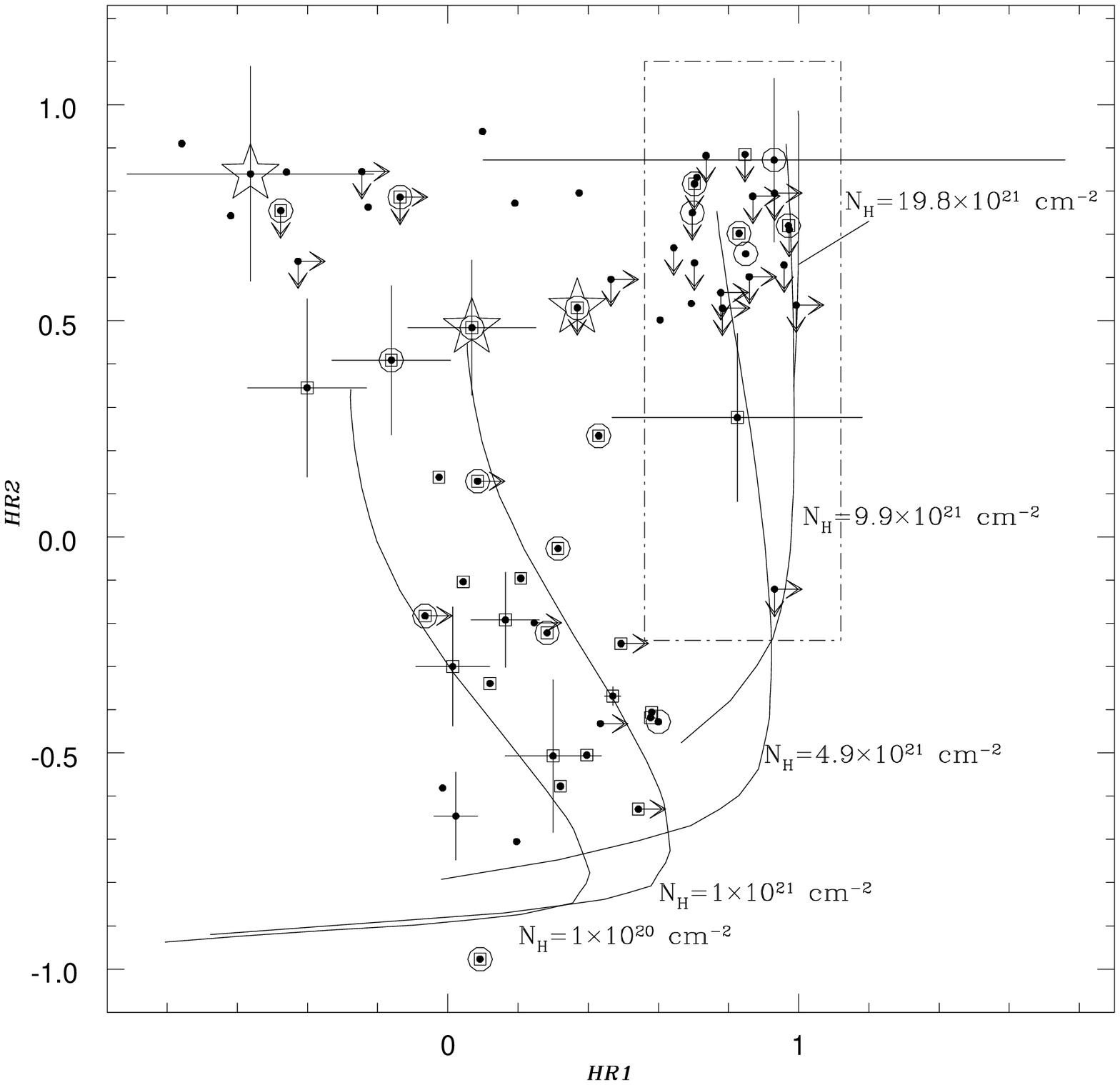} \hspace{4mm}
  \includegraphics[width=8cm]{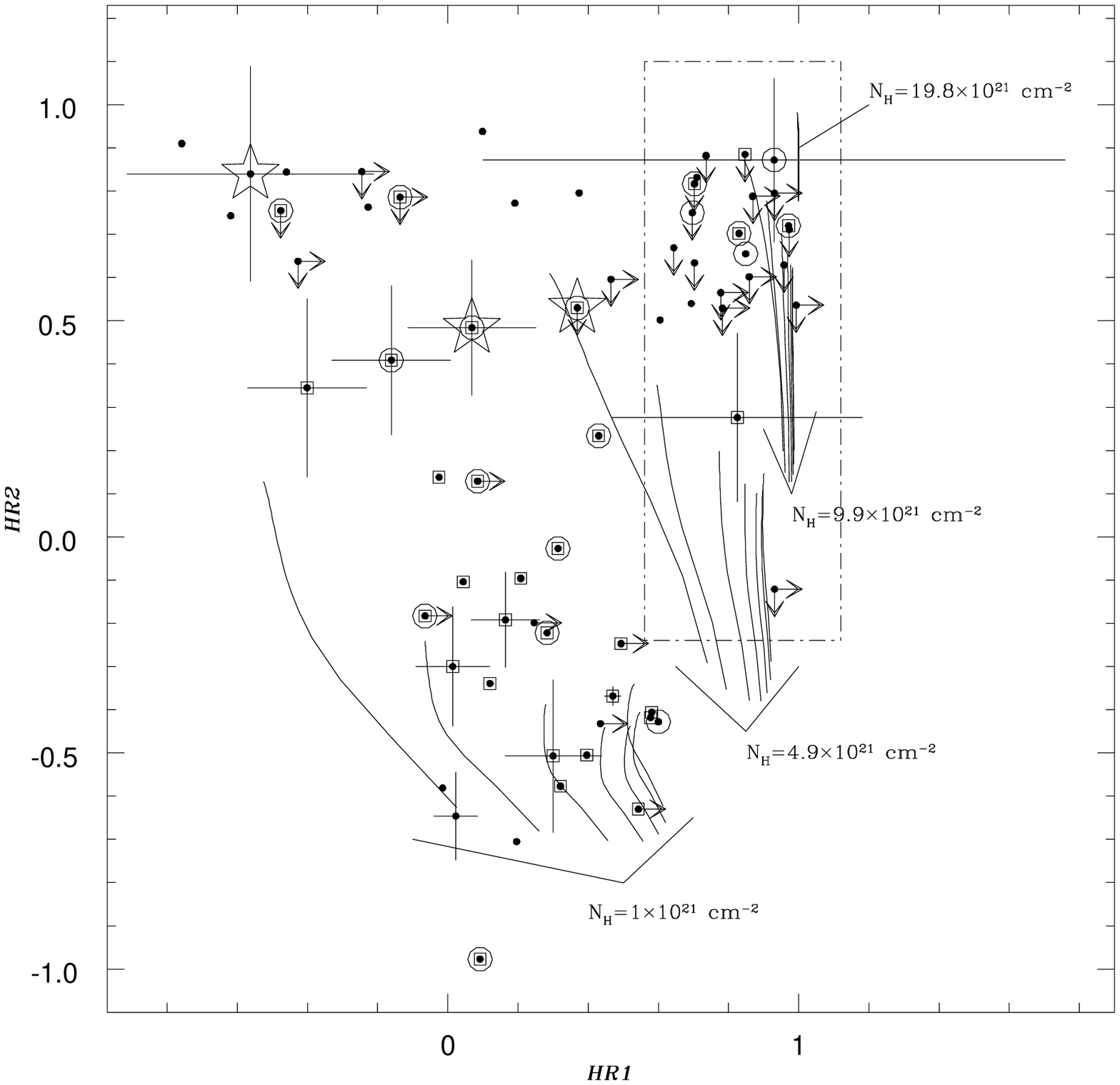}
  \caption{$HR1$ vs $HR2$ diagram for the X-ray sources
    denoted by solid dots. X-ray sources with optical
    counterparts, NIR counterparts and extended behavior are 
    denoted by open circles, open squares and the symbol of star, respectively. 
    X-ray sources with rightward arrow represent the upper limits in $HR1$,
    and downward arrow represent the upper limits in $HR2$.
    Solid lines represents $HR1$ vs $HR2$ curves obtained from 
    APEC model and dotted box contains the X-ray probable 
    members sharing the cluster environment.
    $left panel$  For the single temperature 1T-APEC plasma model, we considered plasma 
    temperatures of kT ranges from 0.2 to 8.0 keV, 
    at different  hydrogen column density ($N_H$)=  $1.0\times10^{20}$,
    $1.0\times10^{21}$, $4.9\times10^{21}$, $9.9\times10^{21}$ (for 
    cluster region), $19.8\times10^{21}$ $\rm{cm^{-2}}$. The temperature 
    increases from bottom to top for each $N_H$ value. 
    $right panel$ The two temperature 2T-APEC plasma model with 
    kT1 = 0.2, 0.3, 0.4, 0.5, 0.6, 0.7, 0.8 keV with the combination 
    of kT2 ranges from 0.9 to 9.6 keV, at different  hydrogen column 
    density ($N_H$)= $1.0\times10^{21}$, $4.9\times10^{21}$, 
    $9.9\times10^{21}$ (for cluster region), 
    $19.8\times10^{21}$ $\rm{cm^{-2}}$. 
    The temperature kT1 increases from left to right for each $N_H$ value 
    with increasing kT2 from bottom to top.}
  \label{fig:hratio}
  \end{figure*}

We have determined the X-ray hardness ratio $HR1$ and $HR2$ defined as

\begin{equation}
HR1= \frac{(M_X - S_X)}{(M_X + S_X)}
\end{equation}

\begin{equation}
HR2= \frac{(H_X - M_X)}{(H_X + M_X)}
\end{equation}

\noindent
where $S_X$, $M_X$ and $H_X$  denote the soft band (0.3 -- 0.7 keV), the medium 
band (0.7 -- 1.2 keV) and the hard band (1.2 -- 7.5 keV), respectively.
For undetected sources, we used upper limits as a function of distance 
from the center of FOV to its edges in each the energy bands using the 
method mentioned in \S\ref{sec:detLimit}.
In cases where no counts are observed in any one energy band, 
the $HR1$ or $HR2$ are either +1.0 (no soft counts) or -1.0 (no hard counts),
we have replaced the zero counts value by the upper limits depending upon the 
source position in the CCD in that energy band.
The hardness ratios $HR1$ and $HR2$ of the X-ray sources in the XMM-Newton 
FOV are displayed in Figure~\ref{fig:hratio}. X-ray sources are represented 
by solid dots, X-ray sources having optical counterparts, NIR counterparts 
and extended behavior are denoted by open circles, open 
squares and the symbol of star, respectively. 
X-ray sources with rightward arrows represent the upper limits in $HR1$,
and downward arrows represent the upper limits in $HR2$.
Solid lines represent hardness ratios derived from model spectra.

We have simulated the values of $HR1$ and $HR2$ using
plasma model APEC (Atomic Plasma Emission Code) to relate 
these hardness ratios to the spectral properties of the sources
and to decide the membership of X-ray sources in the cluster region.
Using the EPIC-pn response matrices 
within XSPEC, we have generated model spectra for monothermal plasma (1T-APEC)
and two-temperature plasma (2T-APEC : the coronal X-ray emission from 
active late-type stars is generally not 
monothermal and consistent with the 2T thermal plasma 
models, \citep[see][]{favata03, tsujimoto02,stelzer06}).
We considered plasma temperatures (kT) from 0.2 keV to 8.0 keV for 1T-APEC model,
and kT1 from 0.2 to 0.8 keV with the combination of kT2 from 0.9 to 9.6 keV
for 2T-APEC model at different  
hydrogen column density ($N_H$)=$1.0\times10^{20}$,
$1.0\times10^{21}$, $4.9\times10^{21}$, $9.9\times10^{21}$ 
(for cluster region), $19.8\times10^{21}$ $\rm{cm^{-2}}$.

The X-ray sources lying in between 
$1\times10^{20}~$ $< N_H < 1\times10^{21}~$$\rm{cm^{-2}}$ 
are probably the foreground stars having either NIR counterparts 
or optical counterparts. As the value of mean $N_H$ for the cluster  
is estimated as $9.9\times10^{21}~$$\rm{cm^{-2}}$ from the optical studies,  
 X-ray stars located in between 
$4.9\times10^{21}~$ $< N_H < 19.6\times10^{21}~$$\rm{cm^{-2}}$
belong to nearly the same cluster environment.
In Figure~\ref{fig:hratio}, the X-ray sources situated inside the dotted 
box are considered as  probable members belonging to the cluster. 
Most of the X-ray sources inside this box are having only upper 
limits in the soft bands and have hard spectra, which is a consequence 
of the absorption 
of the soft components of the energy by the large column densities.
A few sources are fitted neither by 1T-APEC model nor 2T-APEC model
and show a soft energy component as well as a hard energy component.
Most of these sources are unidentified sources not having any 
NIR or optical counterparts. They may be the foreground stars,
whose positions are uncertain in hardness ratio diagram
because of the large uncertainty in the hardness ratios.
But, for a few X-ray sources, the uncertainty   
in hardness ratios is not so large that their
location can be explained within the temperature range of the models.
They may be either the foreground stars having very high hard energy 
component greater than 9 keV or the 1T-APEC and 2T-APEC model are not enough
to characterise them.

\subsection{X-ray Luminosity Function}\label{sec:xraLum}

  \begin{figure}
  \includegraphics[height=8cm, angle=270]{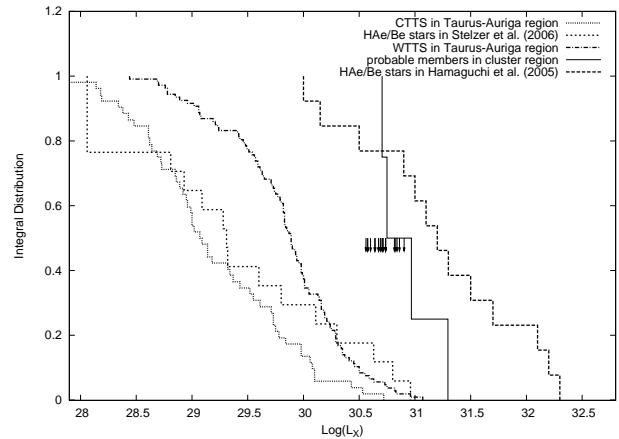}
  \caption{XLF for the probable members in the cluster region NGC 7419, 
    weak-lined (WTTS) and classical T-Tauri stars (CTTS) in Taurus-Auriga 
    region, and Herbig Ae/Be (HAe/Be) stars from \citet{stelzer06} and 
    \citet{hamaguchi05} using Kaplan-Meier estimator. The downward arrows 
    represent 21 Herbig Ae/Be stars in the cluster NGC 7419.}
  \label{fig:xlf}
  \end{figure}

X-ray luminosity function (XLF) is frequently employed to characterize 
a stellar population. We performed  statistical analysis of XLF using 
Kaplan-Meier estimator of integral distribution functions.
We have derived the median $\rm{L_X}$ of weak-lined and 
classical T-Tauri stars in Taurus-Auriga region (age$=0.1-10$ Myrs) using 
the data from \citet{stelzer01}
and for Herbig Ae/Be stars, we have used the sample  
of \citet{stelzer06} and \citet{hamaguchi05}.
The median $\rm{L_X}$ of four probable members in the cluster
is estimated as 30.75 $\rm{erg~s^{-1}}$.
For weak-lined T-Tauri, classical T-Tauri, and Herbig Ae/Be stars
in the sample of \citet{stelzer06} and \citet{hamaguchi05},
the median $\rm{L_X}$ have been estimated as 29.88, 
29.07, 29.29 and 31.15 erg s$^{-1}$, respectively.
A comparison of the XLFs of T-Tauri stars and Herbig Ae/Be stars 
with the cluster members are shown in Figure~\ref{fig:xlf}.

The X-ray upper limits indicate that the Herbig Ae/Be stars
in NGC 7419 are not systematically more active than T-Tauri 
in Taurus-Auriga region and  Herbig Ae/Be stars in \citet{stelzer06}.
But the median value of  XLF of Herbig Ae/Be stars in \citet{hamaguchi05} 
is more than the upper limits in the cluster. 
Our exposure and resolution (see \S\ref{sec:extXra}) are not enough 
to reach the median value of $\rm{L_X}$ of T-Tauri stars, therefore, 
we can not conclude whether Herbig Ae/Be stars can have an XLF
similar to either the T-Tauri stars or Herbig Ae/Be stars in \citet{stelzer06}.
A much deeper detection level is clearly needed in order to reach definitive
conclusions concerning in the X-ray emission from Herbig Ae/Be stars.
However, the more important conclusion to draw from this analysis is that the 
Herbig Ae/Be stars in NGC 7419 are not as active as in the sample of 
\citet{hamaguchi05}.

\subsection{Extended X-ray Emission}\label{sec:extXra}

  \begin{figure}
  \centering
  \includegraphics[width=8cm]{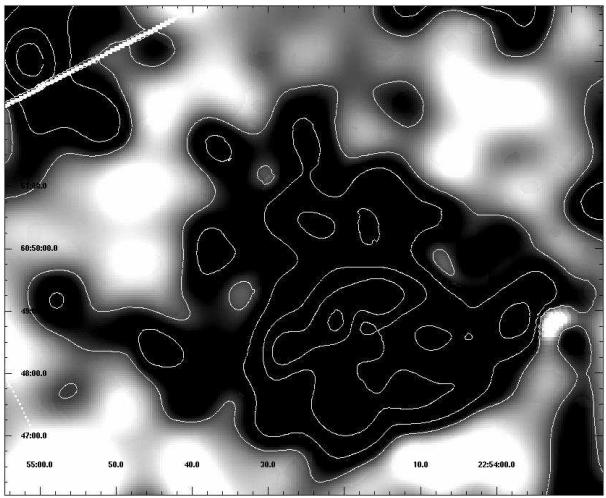}
  \caption{The X-ray extended emission from the cluster NGC 7419 in 
    mosaic MOS detector in energy band the 1.2 -- 7.5 keV.
    X-axis and Y-axis denote $\rm{RA_{J2000}}$ and $\rm{DEC_{J2000}}$, respectively.  
    The gray scale is logarithmic to highlight the diffuse component.
    The contours are  plotted at 4.104 (above $3\sigma$ of 
    the background), 4.644, 5.418, 6.192, 
    6.966 $\rm {10^{-3}~counts~s^{-1}~arcmin^{-2}}$.}
  \label{fig:diffuse}
  \end{figure}

  \begin{figure}
  \centering
  \includegraphics[width=8cm]{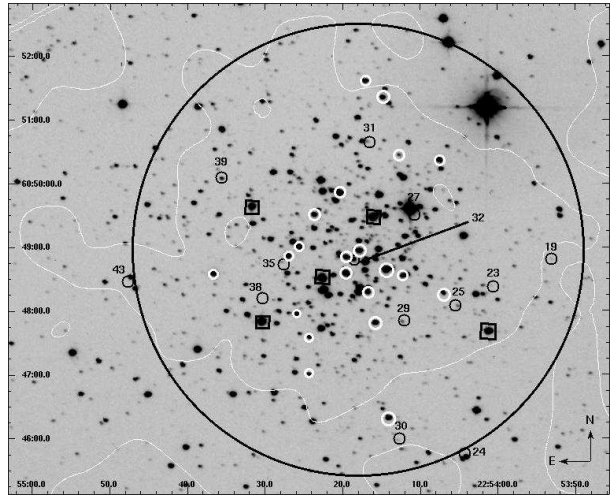}
  \caption{Overlay of X-ray point sources (black open circles) with  
    ID from Table~\ref{tab:xraysrc} and 21 Herbig Ae/Be stars (white 
    open circles) in the cluster region NGC 7419. Large black circle represent 
    the cluster region and black squares represent supergiants.
    X-axis and Y-axis denote $\rm{RA_{J2000}}$ and $\rm{DEC_{J2000}}$, respectively.  
    The white contours represent the extended X-ray emission in energy 
    band 1.2 -- 7.5 keV above the $3\sigma$ of the background count 
    rate i.e. the outermost contour of the Figure~\ref{fig:diffuse}.}
  \label{fig:diffuseOpt}
  \end{figure}

The extended emission from the cluster region NGC 7419 
has been studied using MOS images. PN data were not used 
since chip gaps, bad columns, and the the out-of-time
 event correction would have complicated the analysis.
We have created the MOS images and associated exposure maps in three bands:
soft band (0.3 -- 0.7 keV), medium band (0.7 -- 1.2 keV)
and hard band (1.2 -- 7.5 keV).
We have chosen  adaptive smoothstyle with signal-to-noise ratio 10
in the task $\sc asmooth$, for getting the faint emission from the cluster 
region. The resulting mosaic image with point sources are weighted, masked 
and exposure-corrected following the procedure described in the documentation 
for the SAS task $\sc asmooth$
\footnote{http://xmm.gsfc.nasa.gov/docs/xmm/sas/help/asmooth/index.html}.
The point sources are removed from the image using cheesemask 
but the background is not subtracted from it.

The signature of extended emission is not present in the 
soft band (0.3 -- 0.7 keV) and medium band (0.7 -- 1.2 keV) images of the MOS,
because of the  the absorption 
of the soft X-ray component due to the large hydrogen column density ($N_H$).
All the probable point sources have hard spectra, 
therefore, it is very likely that the extended emission is due to unresolved 
hard X-ray sources in the cluster. The X-ray contour map of the diffuse 
X-ray emission in energy band 1.2 -- 7.5 keV
is shown in Figure~\ref{fig:diffuse}. The entire cluster region is covered 
by the central CCD of the MOS, therefore, we have
chosen background regions of 22745 pixels beyond the diffuse emission but 
within the central CCD. The contours are plotted above the 3$\sigma$ of the 
mean background count rates. The extended X-ray emission from the cluster 
region contains 0.084 $\rm{counts~s^{-1}}$ in 14611 pixels
and corresponds to  5.76$\rm{\times10^{-6}~counts~s^{-1}~pixel^{-1}}$.
The pixel size is set at $2\arcsec\times2\arcsec$.
Therefore, the total extended emission has a size of 16.23 arcmin$^2$. 
We have estimated the count rate from the  extended emission
as 1.38$\rm{\times10^{-3}~counts~s^{-1}~arcmin^{-2}}$, 
after subtracting the mean background count rate.
Using WebPIMMS ver3.9a, this count rate is converted into flux  
after considering Raymond-Smith thermal plasma model at 1 keV
and estimated as $\rm{1.47\times10^{-14}~ erg~s^{-1}~cm^{-2}~arcmin^{-2}}$, 
which corresponds to the 
luminosity $\rm{L_X\approx1.8\times10^{31}~ erg~s^{-1}~arcmin^{-2}}$ in the 
energy band  1.2 -- 7.5 keV. We have  chosen the beamwidth as 50 pixel 
i.e., $5\sigma$ (smoothing beam width). We have estimated 18 beam elements 
in per arcmin$^2$ region and estimated $\rm{L_X}$ of unresolved  
sources as  $\rm{1.0\times10^{30}~ erg~s^{-1}}$. Therefore, we are 
expecting $\sim$ 288 unresolved X-ray sources in the total diffuse component.
 
The $\rm{L_X}$ of  unresolved X-ray sources is comparable with the 
$\rm{L_X}$ of T-Tauri stars 
i.e., 28 $<$ log $L_X$   $<$ 32 in energy band 0.5 to 8.0 keV, with
a peak around log $L_X  \sim $ 29, \citep{feigelson05}. 
It can be explained by the presence
of  $\sim$ 288 T-Tauri stars in the cluster, which are not
resolved by XMM-Newton observatory.
Higher resolution (sub arcsec) data 
are urgently required to explore this issue.

\subsection{X-Ray emission and Herbig Ae/Be stars}\label{sec:xraEmi}

We have examined a sample of 21 Herbig Ae/Be in this cluster
using 2MASS data, \halpha photometry
and the spectroscopic studies reported by \citet{subramaniam06}. 
However, none of Herbig Ae/Be stars has an X-ray counterpart.
In Figure~\ref{fig:diffuseOpt}, the distribution of the 
X-ray point sources which are having optical counterparts, 
supergiants (5 in number) and the Herbig Ae/Be stars within the cluster are 
shown by black open circles (with ID from Table~\ref{tab:xraysrc}), black 
open squares and white open circles, respectively.

The detection limit analysis \S\ref{sec:detLimit} and the 
XLF (Figure~\ref{fig:xlf}) indicate that 
the $\rm{L_X}$ observed from the low mass pre-main-sequence stars
is less than upper limits of the detection 
\citep[see,][]{stelzer01} i.e., 28 $<$ log $L_X$ $<$ 32 in energy 
band 0.5 to 8.0 keV, with 
a peak around log $L_X\sim $ 29  $\rm{erg~s^{-1}}$, (Feigelson et al. 2005).
Therefore, if a Herbig star is 
having a T-Tauri star as a binary companion, then we cannot detect the
X-ray emission from it. Therefore, the generation of X-ray emission 
from Herbig Ae/Be stars \citep[see,][]{zinnecker94,damiani94,stelzer06} might 
be the result of a T-Tauri binary companion
which is not detectable in our study and we can not rule out the companion 
hypothesis for the generation of  X-ray emission from  Herbig Ae/Be stars.
It might happen that the Herbig Ae/Be star itself is emitting X-rays 
but the level of the X-ray emission is less than the detection limit.
A similar kind of process as in T-Tauri stars could then take place in 
the Herbig Ae/Be stars for the generation of X-rays.

\section{Summary and Conclusions} \label{sec:sumCon}

A deep optical $UBVRI$ and narrow band \halpha observations
along with  multi-wavelength archival data 
from the surveys such as 2MASS, MSX, IRAS 
and XMM-Newton are used to understand the global scenario of star formation
and the basic parameters of the cluster NGC 7419. 
XMM-Newton archival data have also been 
used to study the X-ray emission mechanism from the cluster.

The radius of the cluster NGC 7419 has been found to be 4\farcm0$\pm$0\farcm5 using radial 
density profile. The reddening law in the direction of the 
cluster is found to be normal at longer wavelengths but anomalous at shorter
wavelengths. Reddening, $E(B-V)$, is found to be varying between 
1.5 to 1.9 mag with a mean value $\sim 1.7\pm0.2$ mag. The turn-off age and the 
distance of the cluster are estimated to be $22.5\pm2.5$  Myr and
$\rm{3230^{+330}_{-430}}$, respectively.
The MF for the main-sequence stars in the cluster is estimated as having
$ \Gamma =-1.10 \pm 0.19 $ in the mass range $8.6 M\subsun < M
\leq 1.4 M\subsun$, which is a similar to the \citet{salpeter55} value.
Effect of mass segregation is found in the main-sequence stars 
which may be the result of dynamical evolution.

Using the NIR color-color diagram and narrow band $\halpha$
observations, we have identified
21 Herbig Ae/Be in the cluster region with the masses lying 
between 3 to $7 M\subsun$.
The ages of these Herbig Ae/Be  stars
are found to be in the range of $\sim$ 0.3 to 2.0 Myr.
The significant difference between  turn-off age and  turn-on age of
the cluster represents a second episode of
star formation in the cluster.
We have found 90 YSOs having masses
in the range from 0.1 to $2.0 M\subsun$
with the help of NIR color-color diagram around the cluster.
The presence of  such a large number of  NIR excess sources (T-Tauri stars)
shows a recent star formation episode in the surroundings of the
cluster region. Using extinction, dust and $^{12}$CO maps, 
we found that these YSOs are probably associated with a foreground 
star forming region Sh-154 and not related with the cluster region.
We found no obvious trend in spatial distribution of YSOs with $A_V$
and $(H-K)_{\rm excess}$. The dispersion in $A_V$ and $(H-K)_{\rm excess}$
is also very low which indicates that a majority of the YSOs are born at the
same time in the same environment.
Therefore, it is possible that primordial
fragmentation of the cloud may be responsible for the formation of
the low mass stars within this cloud.

We have detected 66 X-ray sources in the observed field by XMM-Newton observatory using 
archival X-ray data.
Out of these sources, 23 are knowm to be the probable members of the cluster
based on analysis of their X-ray colors.
Fifteen X-ray sources are without any optical or NIR counterparts.
These may be the young embedded sources which need to be investigated further.
We have derived the detection limits of X-ray observations 
based on the position of the X-ray source
in the field of view and on its energy spectrum.
Thus, the median value for the detection limit for the 21 Hebig Ae/Be stars
in the field is $\rm{L_X \sim}$$\rm{5.2\times 10^{30}~erg~s^{-1}}$.
We have compared the XLF for the cluster members with the T-Tauri stars 
and Herbig Ae/Be stars.
Because of insufficient exposure and resolution, the sensitivity
was not enough to reach the level of the median X-ray luminosity observed in
T-Tauri stars in Taurus-Auriga region and Herbig Ae/Be stars in
\citet{stelzer06}. Therefore a conclusive comparison of X-ray properties of  
the stars cannot be made. However, the comparison indicates that  
Herbig Ae/Be stars in NGC 7419 tend to be less X-ray luminous than in  
the sample of \citet{hamaguchi05}, which shows that X-ray activity level
of the Herbig Ae/Be stars is not more than in the T-Tauri stars.
Therefore, we can support the binary T-Tauri companion hypothesis
for the generation of  X-rays  in Herbig Ae/Be stars. It is also
possible that a Herbig Ae/Be star is itself emitting X-rays but
the level of the X-ray emission is similar to that of the T-Tauri stars.
The cluster region shows an extended X-ray emission with a total
luminosity estimated to be $\rm{ L_X \approx
1.8\times10^{31}~erg~s^{-1}~arcmin^{-2}} $.
This diffuse emission might be the result of X-ray emission from T-Tauri 
type stars which could not be resolved by XMM-Newton observations.
It requires $\sim$288 T-Tauri stars  each having 
$\rm{L_X}$ $\rm{\sim 1.0\times10^{30}~erg~s^{-1}}$, if it originates
from such stars. High resolution deep observations such as from CHANDRA 
are required for a detailed analysis of this cluster region.


\section*{acknowledgments}
Authors are thankful to the anonymous referee for constructive comments.
This publication makes use of data products from
XMM-Newton archives using the high energy astrophysics science archive research center which is established at Goddard by NASA. We acknowledge 
Dr. Randall Smith from Goddard Space Flight Center NASA and
XMM-Newton Help Desk for their remarkable support in X-ray data analysis.
This research has also made use of data from the Two Micron All Sky Survey, 
which is a joint project of the University of Massachusetts; the 
Infrared Processing and Analysis Center/California Institute of Technology, 
funded by the National Aeronautics and Space Administration and the National 
Science Foundation and VizieR catalogue access tool, 
CDS, Strasbourg, France. One of us (BK) acknowledges support from the Chilean
center of Astrophysics FONDAP No. 15010003.


\end{document}